\documentclass[%
 reprint,
superscriptaddress,
nofootinbib,
 amsmath,amssymb,
 aps,
]{revtex4-1}

\usepackage{color}
\usepackage{amsmath}
\usepackage{amssymb}
\usepackage{graphicx}
\usepackage{dcolumn}
\usepackage{bm}
\usepackage{hyperref}
\usepackage{cleveref}
\usepackage{pbox}
\usepackage[utf8]{inputenc}
\usepackage[]{times}

\newcommand{\cf}{cf.~}
\newcommand{\ie}{i.e.,~}
\newcommand{\eg}{e.g.,~}

\definecolor{orange}{rgb}{1.0, 0.5, 0.0}

\newcommand{\Msol}{M_{\odot}}

\graphicspath{{figs/}}

\begin{document}

\interfootnotelinepenalty=10000
\preprint{APS/123-QED}

\title{A new public code for initial data of unequal-mass, spinning
  compact-object binaries}

\author{L. Jens Papenfort}
\author{Samuel D. Tootle}%
\affiliation{Institut f{\"u}r Theoretische Physik,
  Max-von-Laue-Strasse 1, 60438 Frankfurt, Germany}

\author{Philippe Grandcl\'ement}
\affiliation{Laboratoire Univers et Th\'eories Observatoire de Paris, \\
  Universit\'e PSL, CNRS, Universit\'e de Paris, 92190 Meudon, France
}%

\author{Elias R. Most}
\affiliation{Princeton Center for Theoretical Science, Princeton University, Princeton, NJ 08544, USA}
\affiliation{Princeton Gravity Initiative, Princeton University, Princeton, NJ 08544, USA}
\affiliation{School of Natural Sciences, Institute for Advanced Study, Princeton, NJ 08540, USA}

\author{Luciano Rezzolla}
\affiliation{Institut f{\"u}r Theoretische Physik,
  Max-von-Laue-Strasse 1, 60438 Frankfurt, Germany}
\affiliation{Frankfurt Institute for Advanced Studies,
  Ruth-Moufang-Strasse 1, 60438 Frankfurt, Germany}
\affiliation{School of Mathematics, Trinity College, Dublin 2, Ireland}

\date{\today}

\begin{abstract}
The construction of constraint-satisfying initial data is an essential
element for the numerical exploration of the dynamics of compact-object
binaries. While several codes have been developed over the years to
compute generic quasi-equilibrium configurations of binaries comprising
either two black holes, or two neutron stars, or a black hole and a
neutron star, these codes are often not publicly available or they
provide only a limited capability in terms of mass ratios and spins of
the components in the binary. We here present a new open-source
collection of spectral elliptic solvers that are capable of exploring the
major parameter space of binary black holes (BBHs), binary neutron stars
(BNSs), and mixed binaries of black holes and neutron stars
(BHNSs). Particularly important is the ability of the spectral-solver
library to handle neutron stars that are either irrotational or with an
intrinsic spin angular momentum that is parallel to the orbital one. By
supporting both analytic and tabulated equations of state at zero or
finite temperature, the new infrastructure is particularly geared towards
allowing for the construction of BHNS and BNS binaries. For the latter,
we show that the new solvers are able to reach the most extreme corners in
the physically plausible space of parameters, including extreme mass
ratios and spin asymmetries, thus representing the most extreme BNS
computed to date. Through a systematic series of examples, we demonstrate
that the solvers are able to construct quasi-equilibrium and
eccentricity-reduced initial data for BBHs, BNSs, and BHNSs, achieving
spectral convergence in all cases. Furthermore, using such initial data,
we have carried out evolutions of these systems from the inspiral to after
the merger, obtaining evolutions with eccentricities
$\lesssim 10^{-4}-10^{-3}$, and accurate gravitational waveforms.
\end{abstract}

\maketitle


\section{Introduction}
\label{sec:intro}

In the era of multi-messenger astronomy, precise initial data (ID) for
numerical-relativity simulations is a key ingredient to studying binary
compact object mergers in order to model the observable phenomenon in the
electromagnetic and gravitational-radiation channels. With the detection
of new gravitational-wave sources we have started to obtain a deeper
understanding of the parameter space of compact binary mergers. From the
first detection of a binary neutron star (BNS) merger GW170817
\cite{Abbott2017} and the exceptionally heavy BNS merger GW190425
\cite{Abbott2020}, to the highly asymmetric systems GW190412
\cite{Abbott2020a} and possible black hole neutron star (BHNS) binary
GW190814 \cite{Abbott2020b}, as well as the $150 \, \Msol$ binary black
hole (BBH) merger GW190521 \cite{Abbott2020c}; our understanding of
binary compact-object formation has been confirmed, enriched, and
challenged at the same time. In addition, pulsar observations have lead
to a rich catalogue of observable neutron stars
\cite{Manchester2005-ATNF-Pulsar-Catalog, Lynch2012, Benacquista2013,
  Alsing2017, Tauris2017}. This includes pulsars giving a strong lower
limit on the maximum mass of a neutron star \cite{Antoniadis2013,
  Cromartie2019}, exhibiting extreme rotational frequencies
\cite{Hessels2006}, as well as binary-pulsar systems
\cite{Lattimer2012rev, lorimer:lr} with significant mass asymmetries
\cite{Martinez2015,Lazarus2016,Tauris2019}, and companions with
appreciable spin frequencies \cite{Lyne04,Stovall2018}.

On the theoretical side, increasingly sophisticated parametric studies on
population synthesis and analyses of possible binary-formation channels
show a broad range of resulting binary configurations with respect to the
total mass and mass ratio (see, \eg \cite{Dominik2013, Tauris2017,
  Kruckow2018}). It is also known that the viscosity of nuclear matter
does not suffice to result in tidal locking of inspiraling binary neutron
stars (BNS) \cite{Kochanek92,Bildsten92} -- although bulk-viscous effects
could be important after the merger of a BNS system \cite{Alford2017} --
and that the eccentricity of a binary of compact objects is extremely low
at merger \cite{Kowalska2011}. Furthermore, thanks to the detection of
GW170817, all of these results have been accompanied by a number of
constraints on the equation of state (EOS) of nuclear matter \citep[see,
  \eg][]{Margalit2017, Bauswein2017b, Rezzolla2017, Ruiz2017, Annala2017,
  Radice2017b, Most2018, De2018, Abbott2018b, Montana2018, Raithel2018,
  Tews2018, Malik2018, Koeppel2019, Shibata2019, Nathanail2021}.

The observational evidence of rather extreme configurations of compact
objects\footnote{For an extended discussion on high spin and mass
asymmetry BNS systems see Appendix A of \cite{Dietrich:2015b}.} --
together with the understanding that unequal-mass systems provide better
constraints on the component masses \cite{Rodriguez2013, Most2020c} --
and the constraints on nuclear matter from the first gravitational-wave
detections of BNS mergers, underline the necessity of exploring the edges
of the parameter space. This is especially true for BNS and BHNS binaries
given the degeneracy between tidal and spin effects of the neutron-star
companion on the inspiral waveform \cite{Favata2014, Agathos2015,
  Harry2018, ZhuX2018}. Investigating possible additional observational
channels to discern the exact nature of the given binary is of major
importance in these cases that require the construction of accurate ID
across the whole viable parameter space.

To date, considerable effort has been put towards the underlying
formulation of the equations \cite{Cook2006, Tichy2017} and their
numerical implementation needed to construct state-of-the-art ID solvers
such as \texttt{TwoPunctures} \cite{Ansorg:2004ds,Ansorg:2005bp},
\texttt{SGRID} \cite{Tichy2019, Dietrich:2015b, Tichy06, Tichy:2009,
  Tichy09} for BNS and BBH; using \texttt{BAM} \citep{Brugmann:2008zz,
  Moldenhauer2014, Dietrich:2018bvi} for BNS, BBH, and boson-neutron-star
binaries; \texttt{COCAL} \cite{Tsokaros2012, Tsokaros2015, Tsokaros2016,
  Tsokaros2018, Tsokaros2019} for BNS and BBH; \texttt{Spells}
\cite{Foucart2008, Pfeiffer:2002wt, Pfeiffer:thesis, Tacik:2016zal,
  Tacik15, Tacik16, Ossokine:2015yla, Mroue:2012kv, Lovelace2008c,
  Buchman2012} for BBH, BNS, and BHNS; and the publicly available
spectral solver \texttt{LORENE} \cite{lorene_web, Grandclement06,
  Gourgoulhon01, Taniguchi01, Taniguchi02a,
  Taniguchi02b,Grandclement:2001ed} for BBH, BNS, and BHNS.
Additionally, significant effort has been put into generating binary
compact object ID featuring low orbital eccentricities
\cite{Pfeiffer:2007yz, Husa:2007rh, Buonanno2011, Tichy11b, Puerrer2012,
  Kyutoku2014}, or generalisations to arbitrary eccentricities
\cite{Moldenhauer2014}.

However, publicly available solvers are severely limited in their
capabilities and, even in the case of LORENE, some subsequent
developments are not shared publicly (see, \eg \cite{Kyutoku2014}). Most
notably, there is no open-source code including the treatment of spinning
neutron stars and eccentricity reduction. In addition, there also exists
a portion of the BNS parameter space -- namely, the one considering the
combination of extreme mass ratio and spins for BNS systems -- that has,
to date, not been explored in the context of constraint-satisfying ID.

This work aims to fill this gap by providing an open-source collection of
ID solvers that are capable of exploring the major parameter space of
BBH, BNS and BHNS IDs. In this work we show the ability to construct
\textit{quasi-equilibrium} and \textit{eccentricity-reduced} ID for
\textit{BBH, BNS, and BHNS} utilising the publicly available
\texttt{Kadath}\footnote{https://kadath.obspm.fr/} spectral solver
library\cite{Grandclement09}.

The \texttt{Kadath} library has been chosen since it is a highly
parallelised spectral solver written in \texttt{C++} and designed for
numerical-relativity applications\cite{Grandclement09}. It is equipped
with a layer of abstraction that allows equations to be inserted in a
\LaTeX-like format. In addition to including an array of built-in
operations, user-defined operations can also be written
incorporated into these equation strings. This capability, together with
other ones, allows for readable and extendable source codes.

Overall, with the suite of ID solvers presented here, compact-object
binaries of various type (BBH, BNS and BHNS) can be constructed with mass
ratio $q \neq 1$ and dimensionless spin parameters $\chi_1 \neq \chi_2
\neq 0$. Furthermore, when considering non-vacuum spacetimes, and hence
for BHNS and BNS, we are able to solve the relativistic hydrodynamic
equations utilising tabulated EOSs and obtain spins near their
mass-shedding limit. This is quite an important improvement as many of
the present ID solvers need to make use of piece-wise polytropic fits of
tabulated EOSs when considering unequal-mass binaries.

The paper is organised as follows. In Sec. \ref{sec:init.data}, we will
cover the mathematical framework necessary to obtain accurate ID in
arbitrary, 3+1 split spacetimes, and that is implemented in these
solvers. In Sec. \ref{sec:num.setup} we describe the system of equations
that are solved for each binary type in addition to the iterative scheme
implemented to obtain these IDs. Finally, we present our results in
Sec. \ref{sec:res} for a number of different binaries, followed by a
discussion in Sec. \ref{sec:discussion}.

\section{Mathematical background}
\label{sec:init.data}

Starting with a Lorentzian manifold $\left(\mathcal{M},g\right)$ with the
standard 3+1 split into spatial and temporal parts of the spacetime the
metric takes the form \cite{Alcubierre:2008, Gourgoulhon2012,
  Rezzolla_book:2013}
\begin{equation}
  g_{\mu \nu}dx^\mu dx^\nu = -\alpha^{2} dt^{2} + \gamma_{ij} (dx^{i} +
  \beta^{i} dt)(dx^{j} + \beta^{j} dt)\,,
  \label{eq:3p1_metric}
\end{equation}
introducing the spatial metric, $\gamma_{\mu \nu} = g_{\mu \nu} + n_\mu
n_\nu$, and, consequently, the normal vector, $n_\nu$, to the spacelike
hypersurface, $\Sigma_{t}$, spanned by this construction
\cite{Gourgoulhon2007} as well as the coordinate conditions set by the
lapse, $\alpha$, and shift, $\beta^i$.
In this way the manifold is topologically decomposed into a product space
$\mathcal{M} = \Sigma_{t} \times \mathbb{R}$ parametrized by a time
parameter, $t$.
Under very general conditions this leads to a well-posed formulation of
the Einstein field equations (EFE) as a Cauchy problem \cite{Choquet52,
  Choquet69, Choquet80}. In this way, the Einstein equations are cast in
to a set of ``evolution equations'' (normally written as first-order in
time partial differential equations in hyperbolic form) and a set of
``constraint equations'' (normally written as purely spatial second-order
partial differential equations in elliptic form). A solution to this
latter set is needed to define the ID needed for the evolution of the
spacetime.

More specifically, the projection of the EFE along the normal of
$\Sigma_{t}$ then leads to the so called Hamiltonian and momentum
constraint equations
\begin{align}
  R+K^{2}-K_{ij}K^{ij} &=16 \pi E\,, \label{equ:hamiltonian} && \\
  D_j K^{j}_{\ \, i} - D_{i} K &= 8 \pi j_{i}\,, \label{equ:momentum} &&
\end{align}
with $K_{ij}$ being the extrinsic curvature of $\Sigma_{t}$, $E$ as
$j_{i}$ the temporal-like and spatial projections of the energy-momentum
tensor $T_{\mu \nu}$, and $D_{i}$ the spatial covariant derivative. In
the following sections we will describe our approach to solve these
coupled elliptic partial differential equations in further detail.

\subsection{eXtended Conformal Thin Sandwich Method}
\label{sec:init.data.vac}

The constraint equations \eqref{equ:hamiltonian} and \eqref{equ:momentum}
hide the physical degrees of freedom that one naturally wants to fix in
order to solve for a specific compact-object binary configuration. First
attempts to disentangle such degrees of freedom were made by Lichnerowicz
\cite{Lichnerowicz44} and later extended by York
\cite{York73}. Proceeding with the latter, York introduced a conformally
decomposed thin-sandwich (CTS) approach \cite{York99}, which was then
further adapted to the extended conformal thin-sandwich method (XTCS)
\cite{Pfeiffer:2002iy, Pfeiffer:2005}.

This method combines the conformal decomposition from CTS of the spatial
metric with respect to a background metric $\tilde{\gamma}_{ij}$
\begin{align}
    \gamma_{ij} = \Psi^{4} \tilde{\gamma}_{ij}\,,
\end{align}
and the traceless conformal decomposition of the extrinsic curvature
\begin{align}
    K_{ij} = \Psi^{-2} \hat{A}_{ij} + \frac{1}{3} K \gamma_{ij}\,,
\end{align}
with a modified equation for $\alpha$. The resulting system can be solved
for the conformal factor, $\Psi$, the shift, $\beta^i$, and the lapse,
$\alpha$, given the freely specifiable conformal metric,
$\tilde{\gamma}_{ij}$, and its time derivative, the trace of the
extrinsic curvature, and its time derivative, as well as the matter
sources from the projected energy momentum tensor.

To further simplify the equations, we make some general assumptions
concerning the freely specifiable quantities. First, we restrict the
solutions to a conformally flat metric
\begin{align}
    \gamma_{ij} = \Psi^{4} f_{ij}\,,
\end{align}
where $f_{ij}=\delta_{ij}$ for Cartesian coordinates, but is, in general,
more complex for other coordinates (\eg spherical). Second, we consider a
maximal slicing $K = 0$ of the spacetime. Third, since we are interested
in quasi-equilibrium initial conditions for compact-object binaries for
which circularisation is extremely efficient \cite{Peters:1964}, we
further assume the existence of a helical Killing vector $\xi^{\mu}$
\cite{Bonazzola97,Friedman02,Blackburn92} given by
\begin{align}
  \xi^{\mu} = t^{\mu} = \alpha n^{\mu} + \beta^{\mu}\,,
  \label{equ:tcoord}
\end{align}
in a coordinate system corotating with the binary describing thus a
stationary system.

While not strictly necessary but very much natural, following ansatz
\eqref{equ:tcoord}, we additionally assume that our ID refer to a moment
of time symmetry, thus with a vanishing time derivative of
$\tilde{\gamma}_{ij}$ and $K$. Subsequently, introducing the spatial
covariant derivative of the conformally related spatial metric,
$\tilde{D}_i$, leads to a simplified XCTS system also known as the
Isenberg-Wilson-Mathews approximation \cite{Wilson89}
\begin{align}
    \tilde{D}^2 \Psi &= -\frac{1}{8} \Psi^{-7} \hat{A}_{ij} \hat{A}^{ij}
        - 2 \pi \Psi^5 E, &&
        \label{equ:SYS_conf}\\
    \tilde{D}^2 (\alpha \Psi) &= \frac{7}{8} \alpha \Psi^{-7} \hat{A}_{ij}
        \hat{A}^{ij}
        + 2 \pi \alpha \Psi^5 (E + 2 S),
        \label{equ:SYS_lapse} && \\
    \tilde{D}^2 \beta^i &= -\frac{1}{3} \tilde{D}^i \tilde{D}_j \beta^j +
    2 \hat{A}^{ij} \tilde{D}_j (\alpha \Psi^{-6})
    \nonumber\\ & \hspace{4mm}+ 16 \pi \alpha \Psi^4 j^i\,,
        \label{equ:SYS_shift}&&
\end{align}
constituting a coupled system of elliptic partial differential equations.
It should be noted that this approximation neglects the gravitational
radiation radiated throughout the prior inspiral.

Under these assumptions, the traceless part of the extrinsic curvature is
defined by
\begin{align}
    \hat{A}^{ij} &:= \frac{\Psi^6}{2 \alpha} (\tilde{D}^i \beta^j + \tilde{D}^j
        \beta^i - \frac{2}{3} \tilde{\gamma}^{ij} \tilde{D}_k \beta^k).
        \label{equ:SYS_Aij}&& \\
        &= \frac{\Psi^6}{2 \alpha} (\tilde{\mathbb{L}} \beta)^{ij}
        \label{equ:SYS_Aij_lo} && \\
    \intertext{where $\tilde{\mathbb{L}}$ is the conformal longitudinal operator
such that when acting on a three-vector $v^i$}
  (\tilde{\mathbb{L}} v)^{ij} &:= \tilde{D}^i v^j + \tilde{D}^j v^i
  -\frac{2}{3} \tilde{\gamma}^{ij} \tilde{D}_k v^k\,. &&
\end{align}
The source terms $E$, $S$, and $j^i$ are projections of the
energy-momentum tensor $T^{\mu \nu}$ and thus depend on the exact nature
of the matter or vanish for vacuum spacetimes. These projections will be
discussed in detail in Sec. \ref{sec:init.data.matter}. Finally, to
ensure that the system (\ref{equ:SYS_conf})--(\ref{equ:SYS_shift}) is
well posed, additional boundary conditions must be imposed that will be
discussed in the next sections.

\subsection{Asymptotically Flat Spacetimes}
\label{sec:init.data.vacbc}

For isolated, binary systems of compact objects in quasi-equilibrium, we
enforce that the spacetime will be asymptotically flat at spatial
infinity. Adopting a coordinate system corotating with the binary, this
translates to (see \eg \cite{Gourgoulhon2007})
\begin{align}
    \label{eq:alpha_bc_inf}
    \lim_{r\to \infty}\alpha &= 1\,, & \\
    \label{eq:Psi_bc_inf}
    \lim_{r\to \infty}\Psi &= 1\,, & \\
    \label{eq:shift_bc_inf}
    \lim_{r\to \infty}\beta^i &= \beta^i_{{\rm cor}}\,,
\end{align}
where at large distances the shift is essentially given by the corotating shift
\begin{equation}
  \label{eq:betacor}
  \beta^i_{{\rm cor}} := \xi^i + \dot{a} r^i
= \Omega \partial^i_{\varphi}(\boldsymbol{x}_c) + \dot{a} r^i \,,
\end{equation}
with $\xi^i$ being the spatial part of the helical Killing vector that
describes the approximate stationary rotation in the $\varphi$-direction
of the binary at infinity. The coefficient $\dot{a}$ will appear in an
expansion modelling a finite infall velocity
\cite{Pfeiffer:2007yz,Husa:2007rh,Buonanno2011}, with
$\partial^i_\varphi$ being the standard flat space rotational vector
field around a given center $\boldsymbol{x}_c$. Fine tuning of both
$\Omega$ and $\dot{a}$ provides an effective way to reduce the residual
orbital eccentricity and a detailed description of how this is
implemented in our code is described in Appendix \ref{sec:ecc}.

However, the corotating boundary condition for the shift
\eqref{eq:shift_bc_inf} is numerically infeasible when used as an exact
boundary condition at spatial infinity, where it diverges. We resolve
this by decomposing the shift as
\begin{align}
  \beta^i &= \beta_0^i + \beta^i_{{\rm cor}}\,,
  \label{eq:shift_decomp}
\end{align}
where $\beta_0$ is the part of the shift not involved in the corotation
and sometimes referred to as the ``inertial'' shift. From
Eq. \eqref{eq:shift_decomp}, together with the condition
\eqref{eq:shift_bc_inf}, the boundary condition
\begin{align}
    \lim_{r\to \infty}\beta_0^i &= 0\,,
    \label{eq:residual_shift_bc_inf}
\end{align}
follows trivially. Note that -- in constrast to \eqref{eq:shift_bc_inf}
-- the condition \eqref{eq:residual_shift_bc_inf} is well-defined
numerically. To see how this condition affects Eq. \eqref{equ:SYS_shift}
while already assuming a moment of time symmetry, we can use
Eq. \eqref{equ:SYS_Aij} and a bit of algebra to rewrite
Eq. \eqref{equ:SYS_shift} as
\begin{align}
    2 \alpha \Psi^{-6} \tilde{D}_j \hat{A}^{ij} &= 16 \pi \alpha \Psi^{4} j^i\,.
    \label{equ:SYS_shift_Aij} &&
\end{align}
By using Eq. \eqref{equ:SYS_Aij_lo} and the fact that
$(\tilde{\mathbb{L}} \partial_\varphi)^{ij} = 0$ for a conformally flat
metric \cite{Baumgarte2010, Tichy2017}, both terms in
Eq. \eqref{eq:betacor} vanish on entering
\eqref{equ:SYS_shift_Aij}. Hence, we can write $\beta_0^i$ in
\eqref{equ:SYS_shift} instead of $\beta^i$ and obtain analytically
equivalent solutions related through the decomposition
\eqref{eq:shift_decomp}.

\subsection{Asymptotic quantities}
\label{sec:init.data.infcond}

The total energy contained in a spacetime can be defined through the
integral of the ADM (Arnowitt-Deser-Misner) \cite{Misner73} Hamiltonian
of General Relativity derived from the Hilbert action, leading to an
integral at spatial infinity \cite{Gourgoulhon2007}. This is the
well-known ADM mass $M_{_{\rm ADM}}$. In the case of the asymptotically
flat spacetimes considered here, the terms in the integral drop off
quickly enough and the integral yields a finite value. Further
simplifying the expression by taking advantage of conformal flatness we
ultimately arrive at
\begin{align}
  M_{_{\rm ADM}} &:= - \frac{1}{2 \pi} \int_{S_{\infty}} D^i \Psi \,
  ds_i\,. \label{equ:Madm} &&
\end{align}
Since this is evaluated at spatial infinity with the spacetime being
asymptotically flat, the surface element $ds_i$ is the flat surface
element of the sphere $S_{\infty}$.

Conversely, an alternative way to measure the mass of a
\textit{stationary} spacetime admitting a Killing vector field
$\xi^i_{(t)}$ is the Komar mass $M_{_{\rm K}}$, which, again, is a
surface integral, but can be evaluated anywhere outside the gravitational
sources \cite{Gourgoulhon2007}. Nonetheless, we compute this quantity
again at spatial infinity that, after simplifying the expression for
conformal flatness, gives
\begin{align}
    M_{_{\rm  K}} &:= \frac{1}{4 \pi} \int_{S_{\infty}} n_j \nabla^i
    \xi^j_{(t)} \, ds_i\,. \label{equ:MK:init}
\end{align}
By substituting (\ref{equ:tcoord}) as our Killing vector, we may rewrite
(\ref{equ:MK:init}) as
\begin{align}
    M_{_{\rm K}} &= \frac{1}{4 \pi} \int_{S_{\infty}} D^i \alpha \,
    ds_i\,. \label{equ:MK}&&
\end{align}

Once the ADM mass has been obtained, we quantify the binding energy
between two compact objects in a specific binary configuration by
comparing the total ADM mass of both constituents in isolation $M_{1,2}$
to the ADM mass of the binary system \cite{Baumgarte2010}
\begin{align}
    E_b &= M_{_{\rm ADM}} - M_1 - M_2 =: M_{_{\rm ADM}} -
    M_{\infty}\,.
    \label{equ:binding}&&
\end{align}

Finally, the ADM angular and linear momentum can be computed at spatial
infinity using
\begin{align}
  J_{_{\rm ADM}} &:= \frac{1}{8 \pi} \int_{S_{\infty}} \hat{A}^{ij} \xi_i
  \, ds_j\,, \label{equ:Jadm}&& \\
  P^i_{_{\rm ADM}} &:= \frac{1}{8 \pi}
  \int_{S_{\infty}} \hat{A}^{ij} \, ds_j\,. \label{equ:Padm}&&
\end{align}

\subsection{Quasi-local quantities}
\label{sec:init.data.bhparam}

To fully constrain the system of equations, each compact object must be
constrained by its characteristic parameters such as spin and mass. For a
given compact object, the rotational state is set by the conformal
rotational vector field, $\partial^i_\varphi$, which is centered on the
coordinate center of the compact object, $\boldsymbol{x}_c$,
\begin{align}
  \xi^i_{\rm{(NS,BH)}} := \partial^i_\varphi(\boldsymbol{x}_c)\,.
\label{eq:def_xi}
\end{align}

For a black hole, we can measure these quantities quasi-locally on the
given excision boundary, \ie the horizon \cite{Ashtekar01a, Ashtekar02a,
  Ashtekar03a} (see also \cite{Jaramillo:2011re, Jaramillo:2011rf} for
the possible measurement of radiative degrees of freedom). As a
simplifying assumption we use the black-hole centered rotational vector
field \eqref{eq:def_xi} as the Killing vector field on the black-hole
horizon. Together with the other assumptions and splitting of the
spacetime fields, the quasi-local spin angular momentum is quantified by
\begin{align}
    \mathcal{S} &:= \frac{1}{8 \pi}\int_{S_{\rm BH}} \hat{A}_{ij}
    \xi^i_{\rm BH} dS^j\,,
    \label{equ:qlm:spin}
\end{align}
being a surface integral on the black-hole horizon.

Additionally, the irreducible mass of the black hole (\ie the mass of the
black hole without any angular momentum contribution) is measured by
computing the surface area of the horizon. In the case of conformal
flatness, this calculation is purely a function of the conformal factor
on $S_{_{\rm BH}}$
\begin{align}
    M^2_{{\rm irr}} &:= \frac{1}{16 \pi} \int_{S_{\rm BH}} \Psi^4
    dS\,.
    \label{equ:qlm:Mirr}
\end{align}

With $\mathcal{S}$ and $M_{{\rm irr}}$, the Christodoulou mass $M_{_{\rm
    CH}}$ can be computed, which gives the total mass of the black hole
incorporating the contribution from the spin angular momentum
\begin{align}
  M^2_{_{\rm CH}} &:= M^2_{{\rm irr}} + \frac{\mathcal{S}^2}{4
    M_{{\rm irr}}^2}\,,
  \label{equ:qlm:Mch}
\end{align}
from which the dimensionless spin of the black hole can be defined as
\begin{align}
    \chi & := \frac{\mathcal{S}}{M_{_{\rm CH}}^2}\,. \label{equ:qlm:chi}
\end{align}
Hence, for a BBH system, the total mass at infinite separation is
\begin{align}
    M_{_{\infty,{\rm BBH}}} &:= M_{_{\rm CH,1}} + M_{_{\rm CH,2}}\,, \nonumber
\end{align}
which is measurable quasi-locally even at finite separations and where
$M_{\rm{CH}, (1,2)}$ are the Christodoulou masses of the two black
holes.

For neutron stars we follow a very similar approach. It has been shown in
\cite{Tacik15} that the quasi-local definition of the spin angular
momentum \eqref{equ:qlm:spin} is also applicable -- at least in an
approximate sense -- for a neutron star in a binary system. In this case,
instead of integrating over a horizon, the integration sphere has to be
placed far enough from the neutron-star center so that it contains all of
neutron-star matter. This leads to an approximate but robust measurement
of the quasi-local spin $\mathcal{S}_{_{\rm QL}}$
\begin{align}
    \mathcal{S}_{_{\rm QL}} &:= \frac{1}{8 \pi}\int_{S_{\rm NS}} \hat{A}_{ij}
    \xi^i_{_{\rm NS}} dS^j\,.
    \label{equ:qlm:NSspin}
\end{align}

In contrast to the measurement of the Christodoulou mass $M_{\rm{CH},
  (1,2)}$ on the horizon of a black hole in a binary system, it is not
possible to accurately measure the ADM mass of a single neutron star when
in a binary. Rather, we take as $M_{\rm{ADM}, (1,2)}$ the ADM mass
corresponding to the isolated spinning neutron-star solution having the
same baryonic mass $M_{b}$ and dimensionless spin. This then provides the
best approximation to the asymptotic ADM mass of the binary neutron-star
system as
\begin{align}
    M_{_{\infty,{\rm BNS}}} &:= M_{_{\rm ADM},1} + M_{_{\rm{ADM},2}}\,. \nonumber
\end{align}
The baryonic mass $M_{b}$ of the neutron stars at infinite separation, on
the other hand, is computed as
\begin{align}
    M_{b} &= \int_{V_{\rm NS}} W \rho \Psi^6 dV\,, \label{equ:baryonic-mass}
\end{align}
where $dV$ is the flat-space volume element and $W$ is with the Lorentz
factor [see Eq. \eqref{eq:Lorentzfac} for a definition].

Note, however, that, in analogy with what is done for a quasi-local measure
of the spin, a quasi-local definition of the stellar ADM mass can be
made as \cite{Tichy2019}
\begin{align}
    M_{_{\rm QL}} &:= - \int_{V_{\rm NS}} D_{i} D^{i} \Psi dV\,, \label{equ:adm-mass-ql}
\end{align}
which is a volume integral over a volume $V_{\rm NS}$ enclosing the
neutron-star matter. It has been shown in \cite{Tichy2019} that this
approximate measurement deviates systematically and is not accurate
enough to constrain the dimensionless spin of a star in a binary
precisely. We use it here only to compare to their results in
Sec. \ref{sec:res:bns:comp}. Finally, using Eq. \eqref{equ:qlm:NSspin}
and a robust definition for $M_{_{{\rm ADM}, (1,2)}}$ we can define the
dimensionless spin parameter for each neutron star to be
\begin{align}
    \chi_{\rm{(1,2)}} & := \frac{\mathcal{S}_{\rm{QL, (1,2)}}}{M_{\rm{ADM,
          (1,2)}}^2}\,. \label{equ:qlm:NSchi}
\end{align}

\subsection{Matter sources and hydrostatic equilibrium}
\label{sec:init.data.matter}

The matter content of neutron-star constituents is modeled by a perfect
fluid \cite{Rezzolla_book:2013}
\begin{align}
    T^{\mu \nu} = \left( e + p \right) u^\mu u^\nu + p g^{\mu \nu}\,,
\end{align}
where $e = \rho (1 + \epsilon)$ is the total energy density, $\rho$ is
the rest-mass density, $\epsilon$ the specific internal energy, $p$ the
pressure, and $u^\mu$ the four-velocity of the fluid. The corresponding
source terms entering Eqs. \eqref{equ:SYS_conf}--\eqref{equ:SYS_shift}
are
\begin{align}
    E   &:= \rho h W^2 - p\,, &&\\
    S^j_{\phantom{j} j} &:= 3 p + (E + p) U^2\,, &&\\
    j^i &:= \rho h W^2 U^i\,, &&
\end{align}
where $S^{ij}$ is the fully spatial projection of the energy-momentum
tensor $T^{\mu\nu}$ \cite{Rezzolla_book:2013}, $h := 1 + \epsilon + p /
\rho$ is the relativistic specific enthalpy, and $U^i$ the spatial
projection of the fluid four-velocity, so that the Lorentz factor $W$ is
defined as
\begin{equation}
  \label{eq:Lorentzfac}
  W^2 := (1-U^2)^{-1}\,.
\end{equation}

A general problem with these source terms in combination with a spectral
approach is the explicit appearance of the rest-mass density $\rho$ and
more specifically its behaviour at the stellar surface.  While the limit
of $\rho$ going to zero at the surface can be well captured by adapted
domains fitted to the neutron-star surface (see
Sec. \ref{sec:num.setup.bns}), the very steep drop in magnitude towards
the surface -- especially for very soft EOSs -- poses a challenge to the
spectral expansion, which exhibits oscillations whose amplitude grows
with increasing the number of collocation points. As a result, this
artefact -- which is basically a manifestation of the Gibbs phenomenon --
affects the residuals of the constraint equations and, therefore, can
prevent reaching a fully convergent solution.

Instead of resorting to filtering of the higher-order terms in the
expansion of $\rho$, we transform Eqs. \eqref{equ:SYS_conf},
\eqref{equ:SYS_lapse} and \eqref{equ:SYS_shift} by multiplying them by
the ratio $p/\rho$. This quantity has a well-behaved spectral
representation and shows no oscillating behavior towards the surface,
where it goes to zero for an EOS $p \sim \rho^\alpha$ with $\alpha > 0$.
The resulting system of equations no longer contains explicit occurrences
of the rest-mass density in the source terms and, thus, the residuals of
the equations are left unperturbed. The degeneracy introduced by $p/\rho$
approaching zero towards the surface is fixed by the matching to the
source-free (vacuum) solution of the spacetime.

In addition to being the source terms of the gravitational equations, the
stars have to be in hydrostatic equilibrium. The governing equations are
the local conservation of the energy-momentum tensor $T_{\mu\nu}$, as
well as the conservation of rest-mass
\begin{align}
    \nabla_{\mu} T^{\mu \nu} &= 0\,, \label{eq:tmunu_cons} &&\\
    \nabla_{\mu} \left(\rho u^\mu\right) &= 0\,, \label{eq:dens_cons} &&
\end{align}
where Eq. \eqref{eq:tmunu_cons} gives rise to the relativistic Euler
equation, which, in the limit of an isentropic fluid configuration, reads
\begin{align}
    u^\mu \nabla_\mu \left(h u_\nu\right) + \nabla_\nu h = 0\,. \label{eq:euler_eq}
\end{align}
We note that isentropy is a very reasonable assumption for an
inspiraling cold and unperturbed neutron star.

Introducing now the spatially projected enthalpy current, $\hat{u}_i := h
\gamma_i^\mu u_\mu$, and using the existence of a helical Killing vector,
$\xi^i$, Eq. \eqref{eq:euler_eq} can be rewritten into the purely spatial
equation \cite{Baumgarte2010,Tichy2017}
\begin{align}
    D_i \left(\frac{h \alpha}{W} + \hat{u}_j V^j\right) + V^j \left(D_j
    \hat{u}_i - D_i \hat{u}_j\right) = 0\,, \label{eq:euler_eq_spatial}
\end{align}
with the spatial ``corotating fluid velocity'', $V^i$, defined as
\begin{align}
    V^i := \alpha U^i - \xi^i\,. \label{eq:corot-vel} &&
\end{align}

The isentropic relativistic Euler equation has an exact first integral in
the two notable cases of a corotating or of an irrotational neutron-star
binary in a quasi-circular orbit. In the former case, the spatial
velocity in the corotating frame is $V^i = 0$, while in the latter the
second term in Eq. \eqref{eq:euler_eq_spatial} drops due to the fact that
$\hat{u}_i$ is irrotational and hence its curl is zero by definition. In
practice, for an irrotational binary we introduce a velocity potential
$\phi$ such that $\hat{u}_i = D_i \phi$ \cite{Shibata98,Teukolsky98}, and
thus the last term in Eq. \eqref{eq:euler_eq_spatial} vanishes
identically.

Following the same approach, Eq. \eqref{eq:dens_cons} can also be cast
into a purely spatial equation
\begin{align}
    D_i \left(\rho W V^i \right) = 0\,, \label{eq:dens_cons_spatial}
\end{align}
which, through $V^i$, gives an elliptic equation for the velocity
potential $\phi$. Solving the first integral of
Eq. \eqref{eq:euler_eq_spatial} in the case of a corotating binary -- or
together with the condition \eqref{eq:dens_cons_spatial} in the case of
an irrotational binary -- leads to solutions satisfying hydrostatic
equilibrium.

Note that, as discussed above, the appearance of the rest-mass density
$\rho$ poses a problem for the spectral expansion. Instead of solving
Eq. \eqref{eq:dens_cons_spatial} directly, we recast it in the form
\begin{align}
    \Psi^{6}WV^{i}\tilde{D}_{i}H+\frac{dH}{d\ln\rho}\tilde{D}_{i}(\Psi^{6}WV^{i})
    = 0\,. \label{eq:phi_eq_trans}
\end{align}
After using the conformal decomposition of the spatial metric,
introducing the new quantity $H := \ln h$, and assuming that
$d\ln\rho/dH$ is strictly monotonic, we obtain an additional elliptic
equation with the Laplacian term hidden in the three-divergence
$\tilde{D}_i V^i$. In practice, however, the Laplacian involves only the
derivatives of the velocity potential, $\phi$, which is therefore defined
up to a constant to be fixed explicitly in order to obtain a unique and
bounded elliptic problem.

In Ref. \cite{Tichy11}, Tichy has generalized the irrotational
formulation to uniformly rotating neutron stars in what is referred to as
the constant rotational velocity (or CRV) formalism as a way
to incorporate neutron-star companions with non-negligible spin angular
momentum. In this case, the specific enthalpy current includes a spin
component
\begin{align}
    \hat{u}_i &= D_i \phi + s_i\,, \label{eq:enth:spin_1} && \\ s^i &=
    \omega \xi^i_{\rm NS}\,, \label{eq:enth:spin_2}
\end{align}
where $s_i$ is a rotational vector field centered on the stellar center
utilising Eq. \eqref{eq:def_xi} for the definition of $\xi_{_{\rm NS}}$,
and which represents a uniform rotation contribution to the fluid
velocity parametrized by $\omega$. Note that although the spin velocity
field in \eqref{eq:enth:spin_1} is fully general, we choose
$\xi^i_{\text{NS}}$ in \eqref{eq:enth:spin_2} such that spin and orbital
angular momenta are aligned, which will be the only restriction that we
impose here on our ID models that are otherwise arbitrary.

In this general form, the spatial fluid velocity $U^i$ and Lorentz factor
$W$ become 
\begin{align}
    W^2 &= \frac{\hat{u}^i \hat{u}_i}{h^2} + 1\,, &&\\
    U^i &= \frac{\hat{u}^i}{h W}\,. &&
\end{align}
Furthermore, after neglecting a number of terms in
Eq. \eqref{eq:euler_eq_spatial} on the assumption that they provide
modest contributions given this ansatz (see \cite{Tichy11, Tsokaros2015,
  Tichy2017} for an in depth discussion) it is possible to obtain an
approximate first integral of the type
\begin{align}
    \frac{h \alpha}{W} + D_i \phi V^i = 0\,, \label{eq:first_integral}
\end{align}
which will consequently be employed for both, irrotational and spinning
neutron-star companions.

Finally, to close the aforementioned system for binaries containing
matter sources we need to specify an EOS that relates the thermodynamic
quantities of rest-mass density, $\rho$, pressure, $p$, and internal
energy, $\epsilon$, or, respectively, the relativistic specific enthalpy,
$h$. The infrastructure employed in our code supports both analytic EOSs,
\eg single polytropes and piece-wise polytropes, but also tabulated EOSs
at zero or finite temperature.

\subsection{Black-hole excision boundary conditions}
\label{sec:init.data.bhcond}

When considering black-hole spacetimes, we follow an excision approach
imposing inner boundary conditions on coordinate spheres, namely,
2-spheres corresponding to marginally outer trapped surface (MOTS), and
such that the vector field $k^\mu$ of outgoing null rays on the surface
vanishes on them \cite{Cook:2004kt, Caudill:2006hw}. Translating this to
the conformally flat XTCS fields yields
\begin{align}
    \beta^i |_{S_{\rm BH}} &=\alpha \Psi^{-2} \tilde{s}^i + \omega
    \xi^i_{\rm BH}\,, \label{eq:bh:bc:bet} && \\ \tilde{s}^i
    \tilde{D}_i (\alpha \Psi) |_{S_{\rm BH}} &=
    0\,, \label{eq:bh_bc_lapse_psi} && \\ \tilde{s}^i \tilde{D}_i \Psi
    |_{S_{\rm BH}} &= - \frac{\Psi}{4} \tilde{D}^i \tilde{s}_i -
    \frac{1}{4} \Psi^{-3} \hat{A}_{ij} \tilde{s}^i \tilde{s}^j\,,
\end{align}
where $\tilde{s}^i$ is the conformally normal unit vector on the surface
of the excision sphere $S_{\rm BH}$, which simplifies to the
flat-space normal vector on a coordinate 2-sphere in the case of
conformal flatness considered here. The rotational state of the black
hole is set by the flat space rotational vector field \eqref{eq:def_xi}
centered on the coordinate center of the black hole horizon and parametrized
by the angular frequency parameter $\omega$. It has been shown in
\cite{Cook:2004kt} that the particular choice of the condition
\eqref{eq:bh_bc_lapse_psi}, albeit being arbitrary, has the advantage
that the lapse in the case of non-spherically symmetric solutions is
not fixed explicitly and can thus adapt across the horizon.

\subsection{Neutron-star boundary conditions}
\label{sec:init.data.matter.bc}

While no excised region needs to be introduced in the presence of a
neutron-star companion, and hence there is no requirement for inner
spacetime boundary conditions on the spatial hypersurface, there are
still two boundary conditions that need to be imposed at the stellar
surface. The first one follows from the fact that the stellar mass
distribution in the binary is inevitably deformed due to the tidal
interaction between the compact objects; this is very different from what
happens in the case of a black hole, where the excision surface is
\textit{defined} to be a coordinate sphere of given radius.
As will be described in more detail in Sec. \ref{sec:num.setup}, the
deformation of the star is tracked by a surface-adapting domain
decomposition with the surface defined in general by $\rho \to 0$,
which we translate to the equivalent boundary condition
\begin{align}
    H &= 0\,. \label{eq:bc:H}&&
\end{align}
Secondly, Eq. \eqref{eq:phi_eq_trans} is only valid inside the neutron
star, since it is only defined within the perfect-fluid matter
distribution. Even more important, the second-order term $D_i V^i$
vanishes for $\rho \to 0$, which is readily seen from
\eqref{eq:dens_cons_spatial}. Therefore, in analogy with the
reformulation \eqref{eq:phi_eq_trans}, and exploiting that for finite
derivatives ${dH}/{d\ln\rho} = \rho {dH}/{d\rho} \to 0$ for $\rho \to 0$,
we can make use of the fact that, by definition, $W \neq 0$ and $\Psi
\neq 0$, so that the boundary condition for the elliptic equation
\eqref{eq:phi_eq_trans} can be written as
\begin{align}
    V^i \tilde{D}_i H &= 0\,, &&
\end{align}
on the stellar surface.

\section{Numerical Implementation}
\label{sec:num.setup}

The equations presented in the previous sections constitute a system of
coupled, nonlinear, elliptic partial differential equations. The solvers
employed in this work are codes built around the
\texttt{Kadath}\footnote{https://kadath.obspm.fr/}
library\cite{Grandclement09}. This spectral solver library is publicly
available and uses spectral methods to solve partial differential
equations arising in the context of general relativity and theoretical
physics. A detailed presentation of the library can be found in
Ref. \cite{Grandclement09}. Here, we just recall the basic features and
the additional functionalities that have been added to make this work
possible.

The physical space is divided into several numerical domains. In each of
them, there is a specific mapping from a set of numerical coordinates
(the ones used for the spectral expansion) to the physical ones.
The vicinity of each object is described by three
domains: a nucleus and two spherical-like shells. In the case of a black
hole, the horizon lies at the boundary between the two shells. As in
\cite{LeTiec:2017ebm}, the radius of the boundary is an unknown of the
problem and is found numerically by demanding that the individual mass of
each black hole has a specific value. Note that when considering a system
with larger mass ratios, \ie $q \gg 1$, additional spherical shells need
to be added to the secondary black hole in order to allow for comparable
resolution towards the horizon when compared to the primary black hole. This
is important since, even though a solution can potentially be found for
the system of equations, the majority of the constraint violations can
still accumulate in the vicinity of the horizon of the smaller black hole.

When considering a neutron star, on the other hand, matter occupies the
nucleus and the first shell, so that the surface of the star lies at the
boundary between the two shells. In this case, the shape of the stellar
surface is not know a priori and must be determined numerically by
using the boundary condition \eqref{eq:bc:H}.
The fact that the boundary of the domain is
a variable has to be taken into account properly when solving the
equations. For instance, the physical radius of the stellar domains is no
longer isotropic, but a varying field when expressed in terms of the
numerical coordinates.

\begin{figure}[t]
    \centering
    \includegraphics[width=.495\textwidth,keepaspectratio]{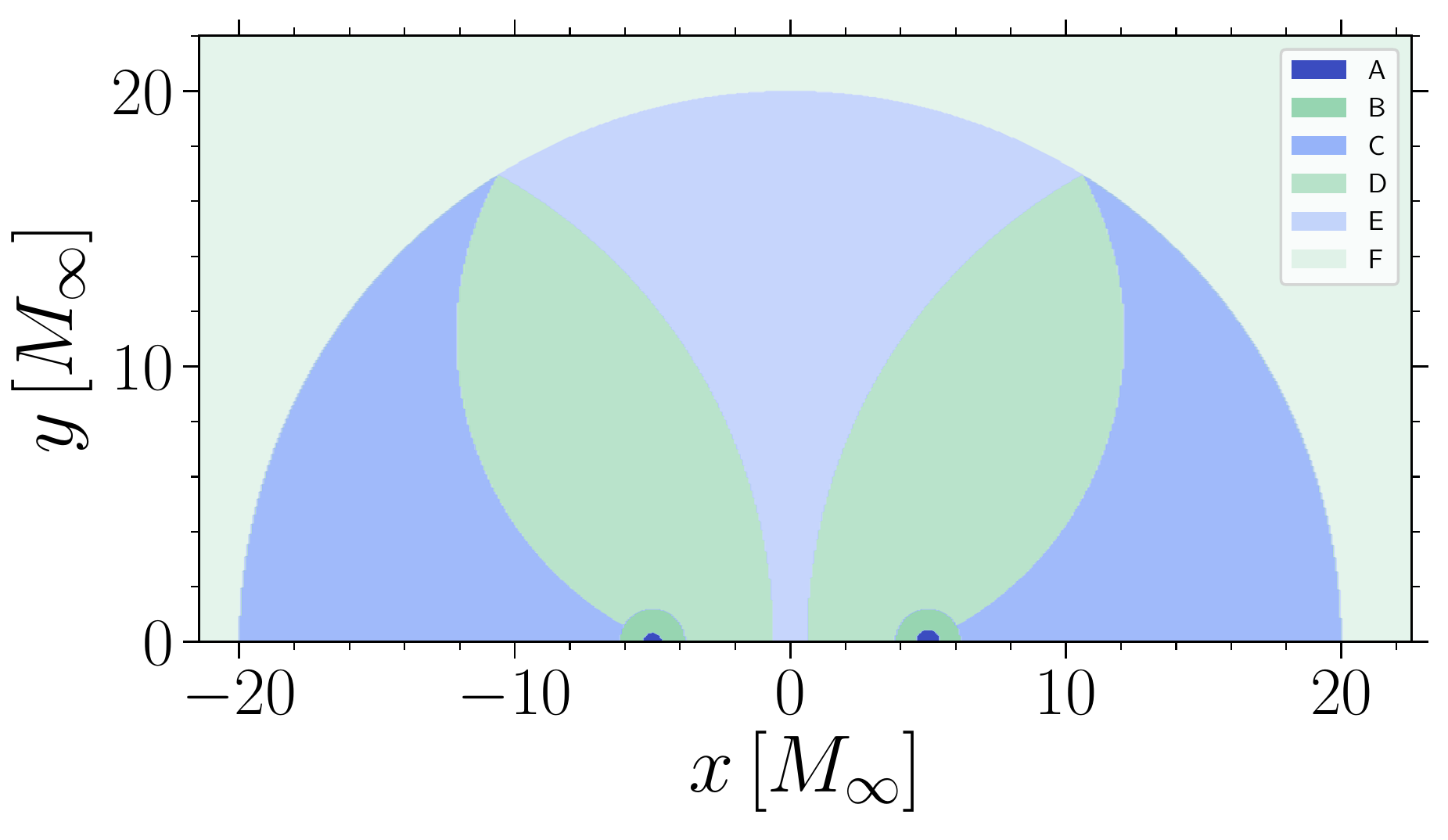}
    \caption{The typical bispherical domain decomposition used in this
      work. Depicted with different shadings are the various coordinate
      domains where: regions A are the excised regions of each BH;
      regions B have a spherical outer radius with an adapted inner
      radius shared A; regions C, D, and E are the bispherical domains,
      and region F is the compactified region. Note that this
      decomposition is rotationally symmetric with respect to the
      $x$-axis. }
    \label{fig:bbh:domain.decomp}
\end{figure}

The connection between the two components of the binary is done via a set
of five domains that implement a bispherical coordinate system. The
description is made complete by an additional compactified domain that
extends up to spatial infinity by means of the use of a compactified
coordinate $1/r$. As a result, the description of a binary system
involves a minimum set of twelve domains. An example of this multi-domain
setting is shown in Fig. \ref{fig:bbh:domain.decomp} where regions A
highlight the excised regions of each BH; regions B have a spherical
outer radius with an adapted inner radius shared with region A; regions
C, D, and E consist of the bispherical domains (see
Ref. \cite{Grandclement09} for their details); and region F is the
compactified region.

In each domain, the fields are described by their spectral expansion with
respect to the numerical coordinates. Chebyshev polynomials are used for
variables with no periodicity, such as the radial coordinate, while
trigonometrical functions are employed for variables that are periodic,
such as the spherical angles of the bispherical coordinates. The choice
of the spectral basis, essentially the parity of the functions, can be
used to enforce additional conditions, such as regularity on an axis of
rotation, or symmetries, like the one with respect to the orbital plane.

Through the spectral representation the residual of the various bulk,
boundary, and matching equations is computed. Depending on the operations
involved, it is more advantageous to represent the fields either by the
coefficients of the spectral expansion or by their values at the
collocation points. Once the residuals are known, they are used to find a
discrete system by means of a weighted residual method. In the case of
the \texttt{Kadath} library, one uses a so-called
``tau-method'', which aims at minimising the coefficients of the
residuals by expanding the residuals $R$ onto a set of test functions
$\xi$ (\ie the domain basis functions) such that the scalar product
$\langle R | \xi\rangle = 0$. In the tau method, the equations
corresponding to the higher order terms can be replaced in order to
enforce boundary and domain matching conditions
\cite{grandclement_2009_smn,Grandclement:2009ju}. The novel parts of the
spectral-solver library introduced in this work refer in particular to
the fluid equations needed when solving for neutron stars, as those
equations have non-standard properties, such as degeneracies at the
surface.  Additionally, modifications of the BBH and BNS spaces along
with the introduction of a BHNS space and major performance optimisation
were essential for this work.

The resulting discretized system is solved by means of a Newton-Raphson
iteration. The computation of the Jacobian of the system is done
numerically and in parallel thanks to the ability of the code to keep
track not only of the value of the fields, but also of their derivative.
This is implemented by the use of automatic differentiation (see Sec. 5 of
\cite{Grandclement09}) and a MPI-parallelised iterative solver.

The equations are, as long as not stated differently, implemented as they
are formulated throughout this paper. By using the capabilities of the
spectral-solver library, the equations are written in a \LaTeX-like format, making
changes and generalisations to the system of equations simple and
straightforward. Since the solution is known as a spectral expansion of
the underlying fields, we generally start generating the solution at very
low resolution with largely reduced computational resources needed for
the first, coarse solution. Interpolating this solution to a space of
higher resolution gives a very good initial guess, so that the Newton-Raphson method
generally converges in only a few steps (down to a single one), depending
on the previous resolution.

The solvers for the different physical binary systems are coded as
stand-alone routines that are linked to the spectral-solver library and
used in conjunction with configuration files in order to steer the
physical parameters, as well as the different solving stages explained in
the next sections \ref{sec:num.imp.bbh}--\ref{sec:num.setup.bhns}.
Additionally, our solvers leverage Kadath's parallel capabilities, which
allows our code to easily scale on high performance computing systems for
an efficient calculation of the ID. As a reference, low-resolution ID
could be obtained within a couple of hours with $\gtrsim 128$ CPU cores, whereas
higher resolution would require $\gtrsim 1000$ and a larger timescale.
Noteworthy the solvers scale almost perfectly with increasing number of
cores up to $\gtrsim 32000$ cores.

\subsection{Binary black-hole (BBH) solver}
\label{sec:num.imp.bbh}

To obtain BBH ID, we employ an iterative scheme that constructs a BBH
system starting from flat-spacetime (\ie $\alpha = \Psi = 1$ and
$\beta_0^i=0$). The system slowly adds constraints over the course of six
stages so as to not introduce too many degrees of freedom initially,
which could result in the solution diverging prematurely. As noted above,
this can be done at very low resolution with only the final step repeated
to obtain the desired final resolution.

In the following we describe the different stages and subsets of
equations that need to be solved numerically to reach a fully constrained
BBH solution.

\subsubsection{Pre-conditioning}
\label{sec:stage.pre}

In the so-called ``pre-conditioning stage'' , we solve only for
Eqs. \eqref{equ:SYS_conf} and \eqref{equ:SYS_lapse}, while enforcing an
initial guess for the fixed radius of the excised region ($R_{\rm BH}=
{\rm const.}$), for a fixed lapse on the horizon ($\alpha|_{S_{\rm BH}} =
{\rm const.}$, where $0 < \alpha < 1$), and a vanishing shift ($\beta_0^i
= 0$). This amounts to solving the Laplace equations for $\alpha$ and
$\alpha\Psi$, and is used to initialise the scalar fields smoothly over
the entire domain decomposition given the inner and outer boundary
conditions before introducing terms involving $\beta^i$.

\subsubsection{Fixed mass and orbital velocity}
\label{sec:stage.fixed.omega}

After the pre-conditioning stage, we solve for the simplest system
involving the shift vector field, which is that of an equal mass,
corotating BBH system with a fixed orbital frequency, namely that given
by a Keplerian estimate obtained using the fixed black-hole masses. Upon
inspection of Eq. \eqref{eq:bh:bc:bet}, it is possible to note that in the
corotating frame the tangential term will vanish when a black hole is
corotating with the binary. Therefore, Eq. \eqref{eq:bh:bc:bet} reduces
to
\begin{align*}
  \beta^i |_{S_{\rm BH}} &=\alpha \Psi^{-2} \tilde{s}^i\,. &&
\end{align*}

In this stage, we solve Eqs. \eqref{equ:SYS_lapse}--\eqref{equ:SYS_shift}
while still utilising a fixed value for the lapse at the boundary of both
black holes ($\alpha |_{S_{\rm BH}}= {\rm const.}$). However, the mass of
the black hole is no longer fixed by a constant radius and, instead, the
variable radius is solved for by imposing a constant irreducible mass
utilising \eqref{equ:qlm:Mirr}.

\subsubsection{Corotating binaries}
\label{sec:stage.fixed.corot}

Next, the same system of equations is solved again, but for a fixed
equal-mass, corotating system, where the orbital angular frequency
$\Omega$ is now fixed by imposing the quasi-equilibrium constraint, \ie
the general-relativistic virial theorem \cite{Gourgoulhon94}
\begin{align}
  M_{_{\rm ADM}} - M_{_{\rm K}} = 0\,.
  \label{equ:virial} 
\end{align}
This results in the first fully self-consistent BBH configuration
representing a corotating black-hole binary in quasi-circular orbit.

\subsubsection{Full system: fixed-lapse boundary conditions}
\label{sec:bbh:total}

Next, the converged corotating solution is further generalized to
arbitrary masses $M_{1,2}$ and dimensionless spins $\chi_{1,2}$ while
still utilising a fixed value of the lapse on the horizon. When obtaining
such solutions there are a few remarks that are worth making.

First, when changing from an equal-mass binary to an unequal-mass binary,
it is important that the total $M_{_{\infty,{\rm BBH}}}$ is kept
constant; failing to do so, \eg allowing for differences in
$M_{_{\infty,{\rm BBH}}}$ as small as $\sim 2\%$, implies that the
solution for the shift from the previous stage will deviate too strongly
from the final result, thereby causing the overall solution to
diverge. Conversely, imposing $M_{_{\infty,{\rm BBH}}} = {\rm const.}$
allows for changes in the mass $q$ up to a factor of $\sim 4$. Second,
using a fixed lapse is essential when solving for a binary for the first
time, or when making significant changes to the parameters of a previous
solution; failing to do so introduces problems in the subsequent stage of
the solver, when von-Neumann boundary conditions are
introduced. Finally, large changes in the mass ratio requires incremental
solutions and, in some cases, higher resolution to obtain a solution to
properly resolve the regions close to the excision boundary.

Note that since, at this stage, the masses are no longer limited to an
equal-mass configuration, the ``center of mass'' of the system is
unconstrained and needs to be determined via the condition that the
asymptotic net linear momentum of the system is zero, \ie
\begin{equation}
  P^i_{_{\rm ADM}} = 0\,.
  \label{eq:centerofmass}
\end{equation}
In practice, since our coordinate system is always centered at the
origin, the corrections coming from enforcing condition
\eqref{eq:centerofmass} -- namely that the helical Killing vector
describes a stationary system that is corotating and centered on the
center of mass -- are added to our helical Killing vector field, which
now takes the form
\begin{align}
  \xi^i = \Omega \partial_{\phi}^i(\boldsymbol{x}_{\rm com})\,,
  \label{equ:com.helical}
\end{align}
where $\boldsymbol{x}_{\rm com}$ represents now the location of the
orbital rotation axis, whose origin we \textit{define} to be the
``center of mass'' of the system in this context throughout this paper.

\subsubsection{Full system: von-Neumann boundary conditions}
\label{sec:bbh:totalbc}

Finally, the von-Neumann boundary condition is imposed on the excision
boundary to relax the necessity to set an arbitrary constant lapse across
the horizon \cite{Cook:2004kt}
\begin{align}
    D_{\boldsymbol{n}} (\alpha \Psi) &= 0\,, &
\end{align}
with $\boldsymbol{n}$ being the normal vector field on the excision
sphere. However, because this boundary condition introduces a considerable
sensitivity to changes in the solution, it is employed only as the final
step of the convergence sequence.

\subsubsection{Eccentricity Reduction}
\label{sec:bbh:ecc}

Strictly speaking the reduction of the eccentricity is not part of the
procedure for finding self-consistent initial data of binary systems,
which completes with the step in Sec. \ref{sec:bbh:totalbc}. Such initial
data, however, although being an accurate solution of the constraint
equations, normally leads to orbital motion that is characterised by a
nonzero degree of eccentricity.
The amount of eccentricity depends sensitively on the
properties of the system (mass ratio and spin) and is most often due to
the various assumptions that are tied with the calculation of the initial
data, \eg quasi-circularity, conformal flatness, etc.

Independently of its origin, such eccentricity represents a nuisance that
needs to be removed as binaries of stellar-mass compact objects are
expected to be quasi-circular in the last stages of the inspiral. In
essence, eccentricity is reduced by utilising input values of $\Omega$
and $\dot{a}$ as constants when solving for the new ID. Since $\Omega$ is
fixed, Eq. \eqref{equ:virial} is neglected in the system of equations to
be solved. Estimates for $\Omega$ and $\dot{a}$ can either be those
derived from approximate treatments, such as post-Newtonian theory [see,
\eg \eqref{equ:Omega3PN} and \eqref{equ:adot3PN} in Appendix
\ref{sec:appendix_PN}] or from an iterative eccentricity reduction
procedure. In this second approach, corrections to $\Omega$ and $\dot{a}$
are calculated by using the ID in short evolutions and by fitting the
orbit using Eqs. \eqref{equ:r-ecc-fit}-\eqref{equ:omega:corr} to obtain
the corrections $\delta \dot{a}$ and $\delta \Omega$ to the previous
estimates \cite{Pfeiffer:2007yz,Husa:2007rh}. The subtleties of this
trial-and-error approach are discussed in detail in Appendix
\ref{sec:ecc} and the included references.

\subsection{Binary neutron-star solver}
\label{sec:num.setup.bns}

When compared to a BBH system, the BNS solver is much less sensitive to
the initial conditions and, therefore, there is no need for additional
sub-stages in the solution process. This is partly due to the fact that
the iterative scheme is started already with a reasonable initial guess
by importing and combining the solutions for static and isolated stars,
(\ie the Tolmann-Oppenheimer-Volkov or TOV equations), but also because
the inner boundary conditions on the excision spheres are susceptible to
small changes in the case of a BBH. In addition, the gravitational fields
and their derivatives are overall smaller and thus the nonlinearities in
the equations less severe.

The scalar fields for the lapse $\alpha_\text{NS}$ and conformal factor
$\Psi_\text{NS}$ from the TOV solutions are interpolated onto the BNS
domains using a simple product of the two independent solutions at a
given Cartesian coordinate $\boldsymbol{x}$
\begin{align}
	\alpha_{_{\rm BNS}} (\boldsymbol{x}) &= 
    \alpha_{_{\rm NS1}}(\boldsymbol{x}-\boldsymbol{x}_{c1}) 
    \alpha_{_{\rm NS2}}(\boldsymbol{x}-\boldsymbol{x}_{c2})\,, &&\\
	\Psi_{_{\rm BNS}} (\boldsymbol{x}) &= 
    \Psi_{_{\rm NS1}}(\boldsymbol{x}-\boldsymbol{x}_{c1}) 
    \Psi_{_{\rm NS2}}(\boldsymbol{x}-\boldsymbol{x}_{c2})\,, &&
\end{align}
where $\boldsymbol{x}-\boldsymbol{x}_c$ represents the coordinate system
with origin in the center of the given compact object.
%
%
Additionally, the matter is imported into the stellar interior domains
and set to zero in all domains outside of the neutron stars.
Given the surface of the stars are described by adapted spherical
domains, the mapping of the adapted domains must also be updated based
on the mappings from the isolated TOV solutions. Finally, the shift is
discarded as the solver is more reliable when starting from zero shift.

As described in Sec. \ref{sec:init.data.matter}, also in the case of a
BNS system we are solving Eqs. \eqref{equ:SYS_conf}--\eqref{equ:SYS_shift}
scaled by the ratio $p/\rho$, together with Eqs. \eqref{eq:first_integral}
and \eqref{eq:phi_eq_trans}, with the additional constraints of $\phi
|_{x_{c}} = 0$ and a fixed $M_{b}$ defined by
Eq. \eqref{equ:baryonic-mass}.

\subsubsection{Full System}
\label{sec:bns:full}

To close the system of equations, there is still the need of a condition
to constrain the orbital frequency, $\Omega$, and, in general, the ``center
of mass'', $\boldsymbol{x}_{\rm com}$. Additionally, the neutron stars --
contrary to a black hole -- have an anisotropic radius distribution along
their adapted surface such that the matter distributions is not
constrained to remain at a fixed distance with respect to the origin of
the innermost domains. To break this degeneracy, we add two conditions
for the two unknowns in terms of the derivative of the enthalpy
\begin{align}
    D_{x} H|_{x_{c_{1,2}}} &= 0\,, \label{eq:force-balance}&
\end{align}
where $x$ is the coordinate direction along which the two stellar centres
are placed and $x_{c_{1,2}}$ are the positions of the fixed centres of
the stars along the $x$-axis. Equations \eqref{eq:force-balance} are the
so-called ``force-balance equations'' \cite{Gourgoulhon01} and complete
the system needed to obtain the ID in quasi-equilibrium.

\subsubsection{Reduced system: fixed linear Momentum}
\label{sec:bns:partial}

In case of high mass ratios, the full system as implemented in stage
\ref{sec:bns:full}, together with Eq. \eqref{eq:force-balance}, yields
binary systems with a non-negligible amount of total linear momentum
$P^i_{_{\rm ADM}}$ at infinity. In turn, this leads to an undesirable
spurious drift of the center of mass of the system during its
evolution. In the same context, we observed that solving
Eq. \eqref{eq:force-balance} separately for each star produces
inconsistent orbital frequencies when considering two stars that differ
significantly in spin and in mass.
Since adding an extra constraint to fix the total
linear momentum renders the system over-determined, and a simple averaging
of the two separate solutions for $\Omega$
\cite{Dietrich:2015b,Tichy2019} is incompatible for the more challenging
configurations involving a high mass ratio combined with extreme rotation
states, we follow a different route.

In particular, we take the matter distribution and the orbital frequency
$\Omega$ computed from the previous stage and define both to be constant,
making Eq. \eqref{eq:force-balance} redundant. At this point, we can use
the condition
\begin{align}
    P^i_{_{\rm ADM}} &= 0\,, \label{eq:lin-mom-constraint}&
\end{align}
to determine a correction to the location of the axis of rotation of the
spatial part of the Killing vector $\xi^i$, just as for a BBH
configuration.

Doing so necessarily leads to slight differences in the velocity field of
the neutron stars due to changes in the velocity potential, which
incorporates and adapts to the different velocity contributions.
Most importantly, doing so introduces small deviations in $M_{b}$
through the Lorentz factor $W$ present in the integral
\eqref{equ:baryonic-mass}\footnote{We note this is true for any solution
with a preassigned $\Omega$, \eg when implementing the iterative
eccentricity reduction discussed in Sec. \ref{sec:ecc}.}. Since the
rest-mass is a fundamental property of the binary from and is conserved
throughout the evolution by \eqref{eq:dens_cons}, it is important to
enforce that the desired value is specified with precision. We accomplish
this by a simple rescaling of the form
\begin{align}
  H \to H &= H_{\rm const.} (1 + \Delta_H)\,,
  \label{eq:matter-rescaling}&
\end{align}
where $H_{\rm const.}$ is the fixed matter distribution from the previous
stage and $\Delta_H$ is the (small) correction needed to enforce that
the baryon mass is the one expressed by Eq. \eqref{equ:baryonic-mass}.

Ultimately, the first integral Eq. \eqref{eq:first_integral} is the only
equation that is violated by the rescaling discussed above, although only
to a limited extent. While this violation certainly has an impact on the
equilibrium of the two stars, this impact is overall negligible. Indeed,
numerical-evolution tests spanning throughout the allowed parameter space
in terms of mass ratio and spin has shown that the perturbations of the
stars are increased insignificantly when compared to the fully
self-consistent solutions resulting from stage \ref{sec:bns:full}.
Furthermore, these perturbations are a priori indistinguishable from those
introduced in the binary simply because of the approximate nature of the
condition Eq. \eqref{eq:first_integral} in the case of high spins. 
More importantly, numerical evolution of high spin and mass ratio
systems without explicitly fixing \eqref{eq:lin-mom-constraint} by using
only \ref{sec:bns:full} exhibit the same orbital evolution as the fixed
systems, apart from a strong center of mass drift.
Thus, the prescription above allows us to have a precise
control of the drift of the center of mass and of the baryon mass of the
binary.

\subsubsection{Eccentricity Reduction}
\label{sec:bns:ecc}

As in the BBH case, in order to reach a solution with reduced orbital
eccentricity, the quantities $\Omega$ and $\dot{a}$ need to be fixed via
an iterative procedure fitting the trajectories in terms of
Eqs. \eqref{equ:adot:corr} and \eqref{equ:omega:corr} so as to obtain the
intended corrections. Since in this case $\Omega$ \textit{has} to be
fixed, we follow the same approach as in stage \ref{sec:bns:partial} and
rescale the matter of the original solution from the stage
\ref{sec:bns:full}. In this way, employing
Eq. \eqref{eq:lin-mom-constraint}, the ID features both a reduced orbital
eccentricity and a very small center-of-mass drift.

\subsection{Black-hole neutron-star solver}
\label{sec:num.setup.bhns}

Finally, to show the flexibility in applying the extended \texttt{Kadath}
spectral-solver library, we can use much of the infrastructure presented above for BBHs
(\cref{sec:num.imp.bbh}) and BNSs (\cref{sec:num.setup.bns}) to construct
binaries composed of a black hole and a neutron star.

For the initial guess, we currently start with an irrotational,
equal-mass system utilising a previously solved BNS ID and an isolated
black-hole solution. This provides a very good estimate for $\alpha$ and
$\Psi$, as well as for the matter-related quantities $\phi$ and $H$. The
shift vector, however, is discarded as this can have a negative impact on
the initial convergence. We note that, in principle, it is also possible
to start directly from a static TOV and single black-hole initial guess
for the spacetime. During the import, spherical shell domains are added
outside of the black hole to obtain the same resolution near the excision
boundary as that which is near the surface of the neutron star. These
additional shells can be removed or added as necessary to obtain the
desired resolution.

\subsubsection{Initial system: fixed-lapse boundary condition}

By combining the two converged datasets, we start our two-stage solver
starting with an initial equal-mass and irrotational BHNS system. More
specifically, in the first stage we solve the neutron-star part using the
same system of equations for the matter component described in
Sec. \ref{sec:bns:full}. On the other hand, when considering the
black-hole component, we utilise the system of equations described
Sec. in \ref{sec:bbh:total}, which fixes the lapse function on the
horizon based on the imported isolated black-hole solution. This stage
proved necessary as the von-Neumann boundary condition was excessively
sensitive and would otherwise result in a diverging solution.

The orbital frequency $\Omega$ of the binary is set solely by
Eq. \eqref{eq:force-balance}, but, unlike for a BNS system where
Eq. \eqref{eq:force-balance} consists of two distinct equations, we still
fix the center of mass by imposing $P^i_{_{\rm ADM}} = 0$ without
over-determining the system of equations.

\subsubsection{Full system: von-Neumann boundary condition}

In this second stage, we repeat the steps just described above, but
exchange the constant lapse constraint on the horizon with the
von-Neumann boundary condition as described in
Sec. \ref{sec:bbh:totalbc}. Once a first configuration has converged
in this stage, all further modifications, such as iterative changes to
the spins and the mass ratio, can be made while subsequently resorting
only to this final stage.

\subsubsection{Eccentricity Reduction}
\label{sec:bhns:ecc}

As in the BBH and BNS scenarios, $\Omega$ and $\dot{a}$ are corrected to
remove the spurious eccentricity by using the same iterative procedures
already mentioned in Secs. \ref{sec:bbh:ecc} and
\ref{sec:bns:ecc}. Additionally, the matter is rescaled as discussed in
\cref{sec:bns:ecc,sec:bns:partial} since $\Omega$ is again a fixed
quantity at this stage. Finally, we explicitly enforce
Eq. \eqref{eq:lin-mom-constraint} to minimised the residual drifts of the
center of mass.

\section{Results}
\label{sec:res}

In the following, we present a collection of ID configurations generated
using the procedures described in Sec. \ref{sec:num.setup}. Such ID is
then employed to carry out evolutions of the various binary systems
making use of the general-relativistic magnetohydrodynamics code
\texttt{FIL} \cite{Most2019b,Most2018b}, which is derived from the
\texttt{IllinoisGRMHD} code \citep{Etienne2015}, but implements
high-order (fourth) conservative finite-difference methods
\citep{DelZanna2007} and can handle temperature and electron-fraction
dependent equations of state (EOSs). Neutrino cooling and weak
interactions are included in the form of a neutrino leakage scheme
\citep{Ruffert96b, Rosswog:2003b, Galeazzi2013}. 

\texttt{FIL} makes use of the \texttt{Einstein Toolkit} infrastructure
\cite{loeffler_2011_et}. This includes the use of the fixed-mesh
box-in-box refinement driver \texttt{Carpet} \cite{Schnetter-etal-03b},
the apparent horizon finder \texttt{AHFinderDirect}
\cite{Thornburg2003:AH-finding} together with \texttt{QuasiLocalMeasures}
\cite{Schnetter-Krishnan-Beyer-2006} to measure quasi-local horizon
quantities of the black holes. The spacetime evolution is done either by
\texttt{McLachlan} \cite{Brown2007b, mclachlanweb} for the BSSNOK
formulation \cite{Baumgarte99,Shibata95} or by \texttt{Antelope}
\cite{Most2019b} implementing the BSSNOK \cite{Baumgarte99,Shibata95}, Z4c
\cite{Bernuzzi:2009ex}, and CCZ4 \cite{Alic:2011a,Alic2013} formulations.

\subsection{Sequences of compact binaries}
\label{sec:res:seqs}

\begin{figure}
	\includegraphics[width=0.475\textwidth,keepaspectratio]{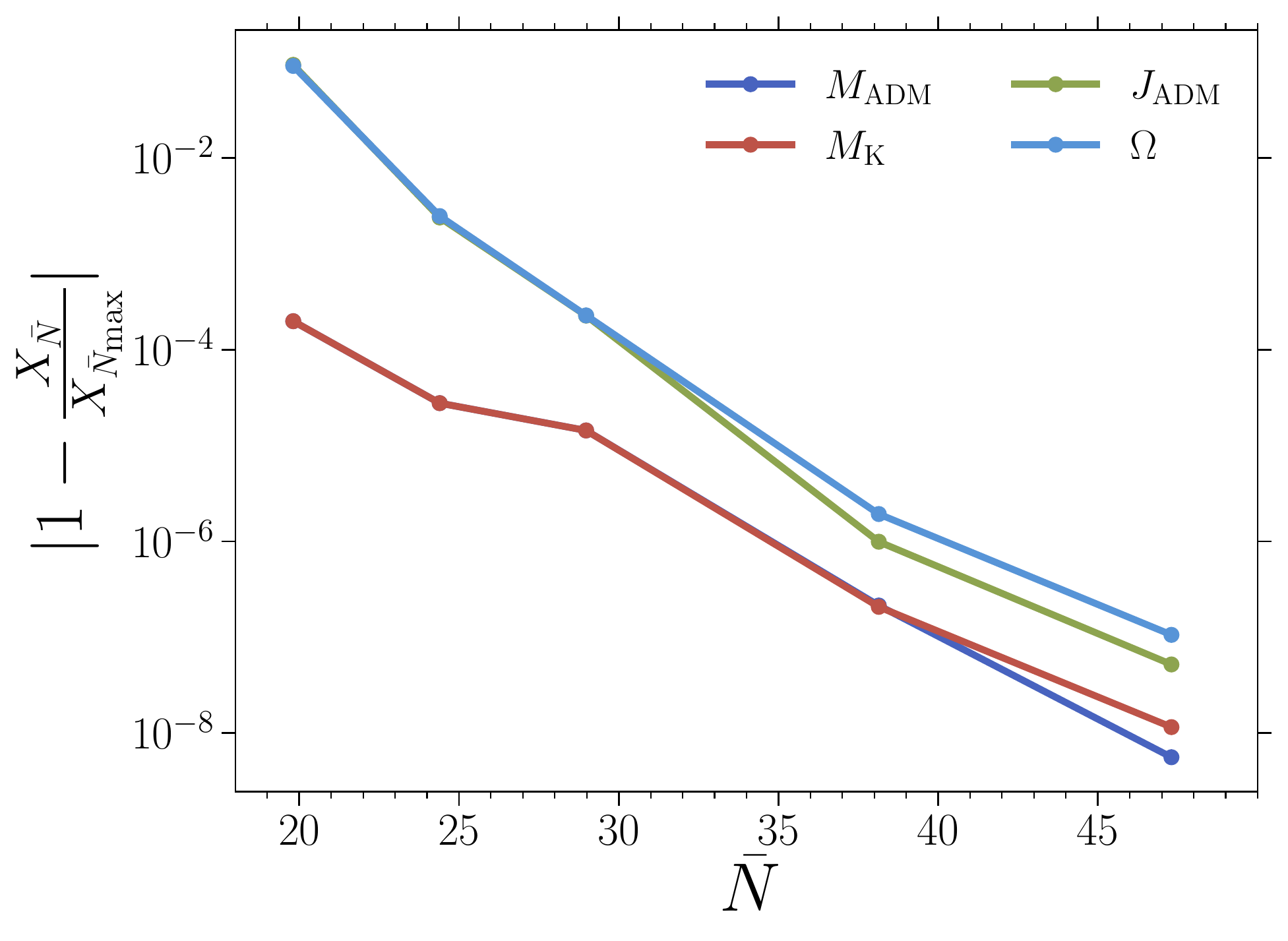}
	\caption{Spectral convergence of the asymptotic quantities
          described in Sec. \ref{sec:init.data.infcond} and of the
          orbital frequency $\Omega$ for an equal-mass BBH system. Shown
          are the absolute values of the variations of the quantity $X$
          at a given effective resolution $\bar{N}$ given by
          Eq. \eqref{eq:eff_resol} with respect to the corresponding
          quantity at the largest effective resolution $\bar{N}_{\rm max}
          = 52$. Clearly the variations decrease exponentially for all
          quantities considered.}
    \label{fig:bbh:convergence}
\end{figure}

As a first result, and as an effective way to quantify the reliability of
our implementations, we perform an initial resolution study to determine
if the global properties of the solutions show the expected spectral (\ie
exponential) convergence for increasing number of collocation points. To
do so we utilise the asymptotic quantities $M_{_{\rm ADM}}$, $M_{_{\rm K}}$,
$J_{_{\rm ADM}}$ defined in Sec. \ref{sec:init.data.infcond} and the
orbital angular velocity $\Omega$ of an equal-mass BBH system. In this
context, we define an effective resolution across the whole space
following \cite{Tacik15}
\begin{align}
    \bar{N} &:= \left( \sum_{i \in \mathcal{D}} N_{\left(i\right)}
    \right)^{\frac{1}{3}}\,, \label{eq:eff_resol}&
\end{align}
where $N_{\left(i\right)}$ is the total number of points of the $i$-th
domain part of the space decomposition $\mathcal{D}$, which is rounded to
the closest integer number. In Fig. \ref{fig:bbh:convergence} we report
for each quantity $X$ (\ie $M_{_{\rm ADM}}, M_{_{\rm K}}, J_{_{\rm ADM}}$,
and $\Omega$), the absolute value of the variations of $X$ at
a given $\bar{N}$ with respect to the corresponding quantity at the
largest value $\bar{N}_{\rm max}$ (\ie the high-resolution solution).
While $M_{_{\rm ADM}}$ and $M_{_{\rm K}}$ have consistently smaller
relative deviations than $J_{_{\rm ADM}}$, and $\Omega$, all quantities
clearly exhibit the expected spectral convergence.

Next, we present quasi-equilibrium sequences of irrotational BBHs, BNSs,
and BHNSs, and compare the corresponding binding energies and orbital
angular velocities with the values obtained from fourth-order
post-Newtonian (4PN) expressions, namely, Eqs. \eqref{equ:EB3PN} and
\eqref{equ:EB4PN} (see, Appendix \ref{sec:appendix_PN} and
\citep{Blanchet2014} for a review).

\begin{figure}
  \includegraphics[width=0.475\textwidth,keepaspectratio]{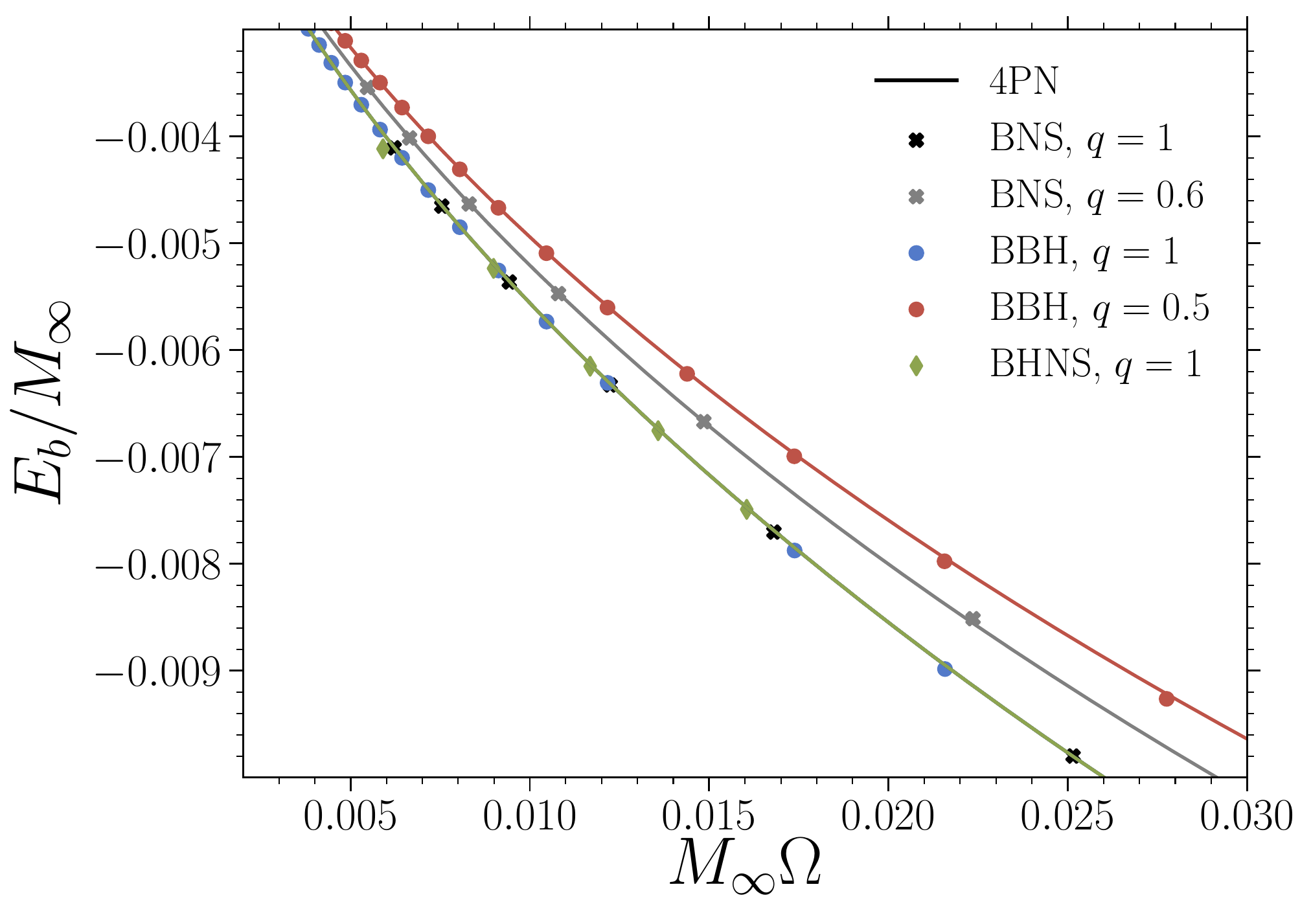}
  \caption{Behaviour of the binding energy [\cf Eq. \eqref{equ:binding}]
    as a function of the dimensionless orbital frequency for sequences of
    irrotational BBH, BNS, and BHNS binaries (coloured filled circles),
    when compared with the combined with the corresponding 4PN prediction
    given by Eq. \eqref{equ:EB4PN} (coloured solid lines). For binaries
    having the same components, we have considered both equal-mass
    binaries ($q=1$) and unequal-mass binaries ($q=0.5$ for BBHs and
    $q=0.6$ for BNSs). Finally, in the case of the BHNS binaries, we have
    computed a rather extreme equal-mass configurations.}
    \label{fig:bbh:binding}
\end{figure}

Figure \ref{fig:bbh:binding}, in particular, presents a comparison of the
binding energy $E_b$ [\cf Eq. \cref{equ:binding}] of various irrotational
compact binaries, namely, BNS (crosses), BBH (filled circles), and BHNS
(diamonds), that have either equal masses ($q=1$) or unequal masses
($q=0.5, 0.6$). In the case of binaries with at least one neutron star,
we model the latter by a single polytrope with $K=100$ and $\Gamma = 2$
as a function of the dimensionless orbital frequency $M_{\infty}
\Omega$. Note that both for equal-mass and unequal-mass binaries our
numerical solutions closely follow the analytical 4PN estimates (solid
lines).

\begin{figure}[t]
  \includegraphics[width=0.475\textwidth,keepaspectratio]{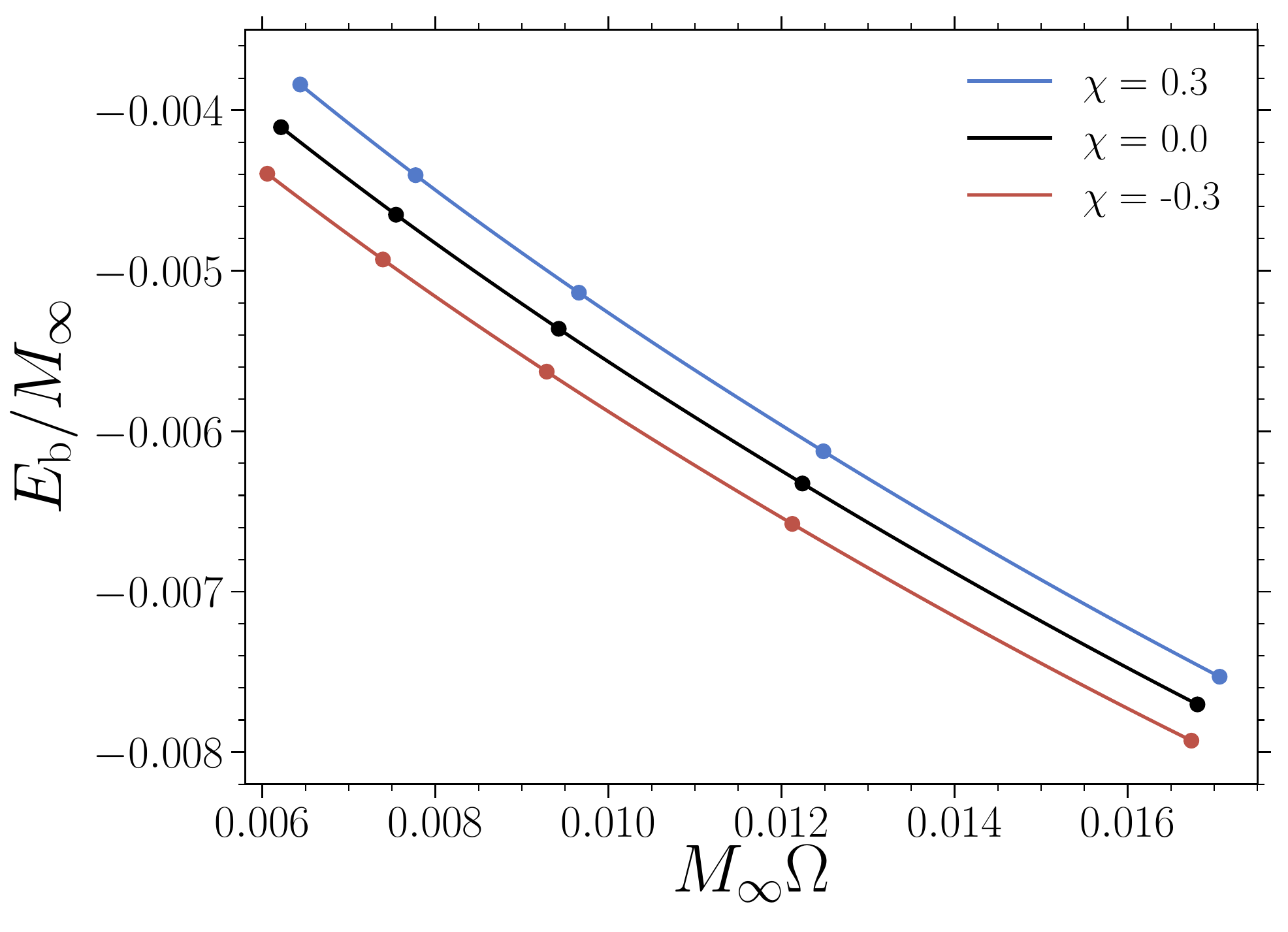}
  \caption{Same as in Fig. \ref{fig:bbh:binding} but for equal-mass, BNS
    configurations that are either irrotational ($\chi=0$) or spinning
    ($\chi=\pm 0.3$), with spins aligned to the orbital angular
    momentum. Also in this case, the coloured solid lines refer to the
    4PN predictions \eqref{equ:EB4PN}, which provide an accurate estimate
    even in the presence of high spins in the range of the given orbital
    frequencies.}
    \label{fig:bns:chi-eb}
\end{figure}

Following a similar spirit, Fig. \ref{fig:bns:chi-eb} reports the binding
energy as a function of the normalised orbital frequency for a selection
of equal-mass, irrotational or spinning BNS configurations with spins
that are either aligned and anti-aligned to the orbital angular
momentum. The EOS used is the same as in Fig. \ref{fig:bbh:binding} (a
single polytrope with $K=100$ and $\Gamma = 2$). As can be clearly seen,
binaries with spin that are aligned with respect to the orbital angular
momentum are less bound than the irrotational counterparts, which, in
turn, are less bound than the binaries with anti-aligned spins. This
result, which is embodied already in the PN equations (see solid lines),
confirms what has been presented in Refs. \cite{Dietrich:2015b,
  Tichy2019} and highlights that binary systems with significant aligned
spins will require a larger number of orbits before merging.

\subsection{Evolutions of black-hole binaries}
\label{sec:res:bbh}

In the following we present the results of the evolutions of BBHs whose
ID have been produced with our new spectral solver. Note that our
evolutions, although comprehensive of all the relevant cases, do not
explore any new aspect of the dynamics of compact binaries that has not
been presented already in the literature. Rather, here they are meant to
be used mostly as representative test cases and clear proofs of the
considerable capabilities of the new spectral-solver library.

\begin{figure*}[t]
  \includegraphics[height=.340\textwidth,keepaspectratio]{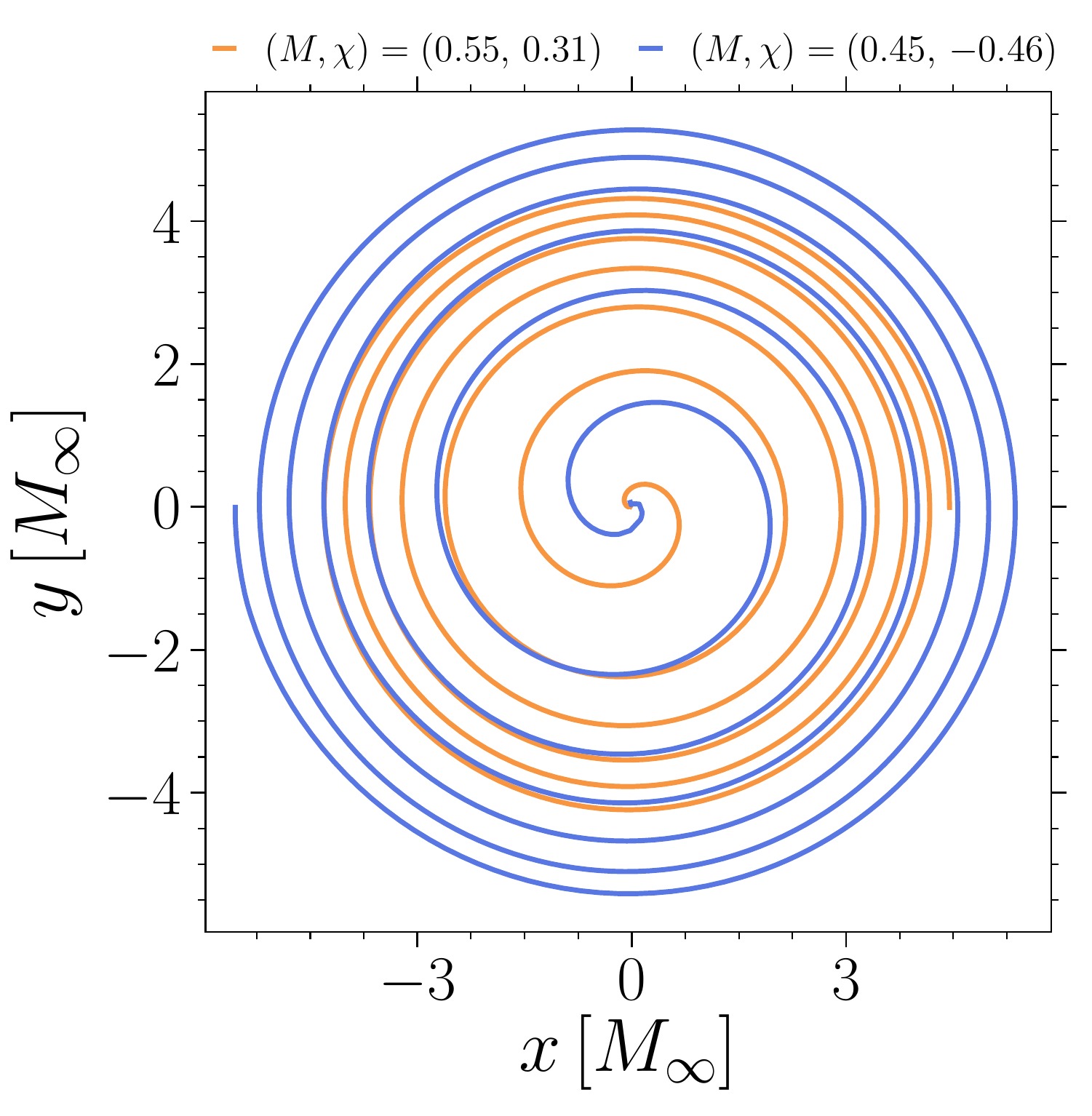}
  \includegraphics[height=.322\textwidth,keepaspectratio]{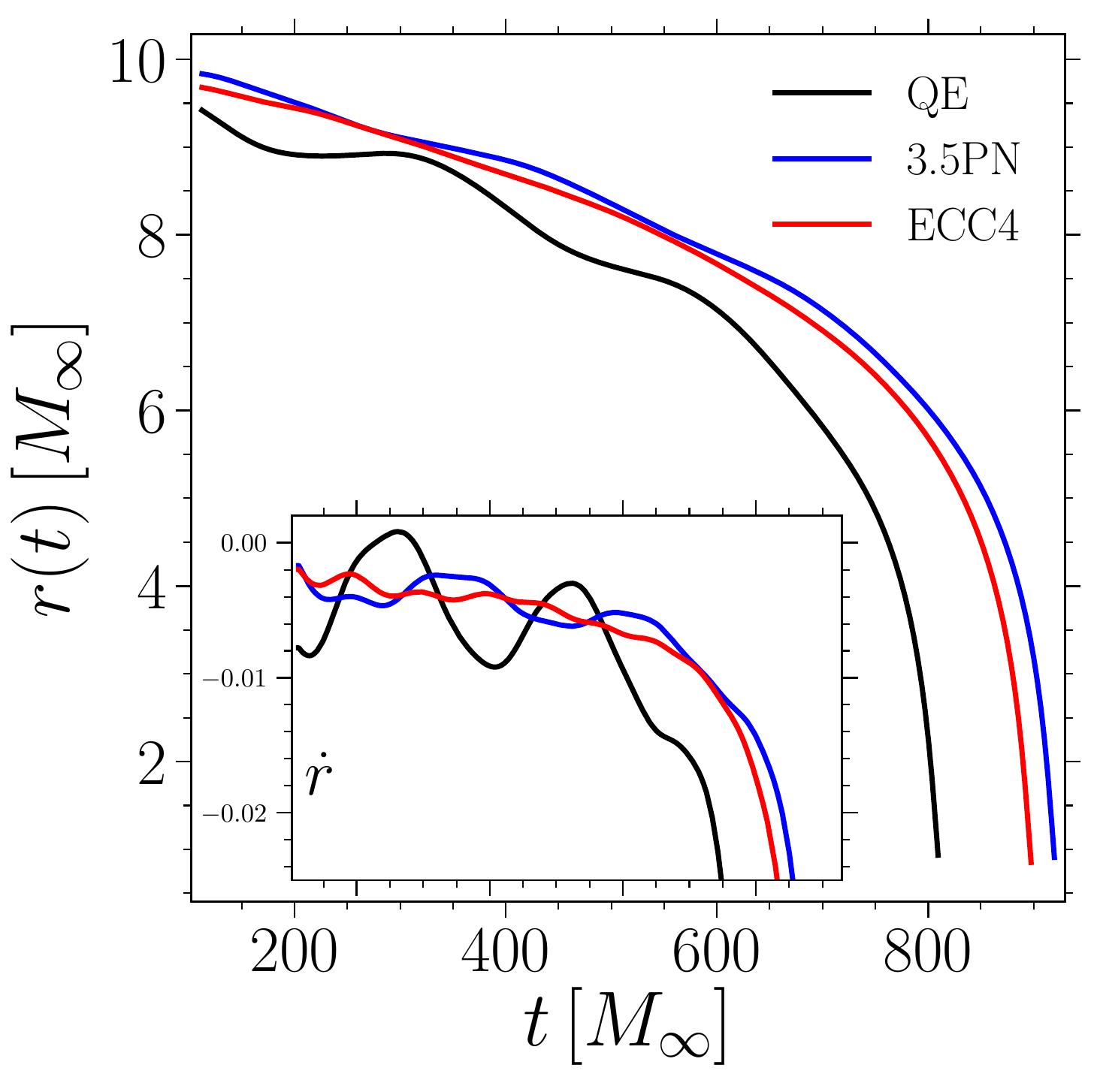}
  \includegraphics[height=.322\textwidth,keepaspectratio]{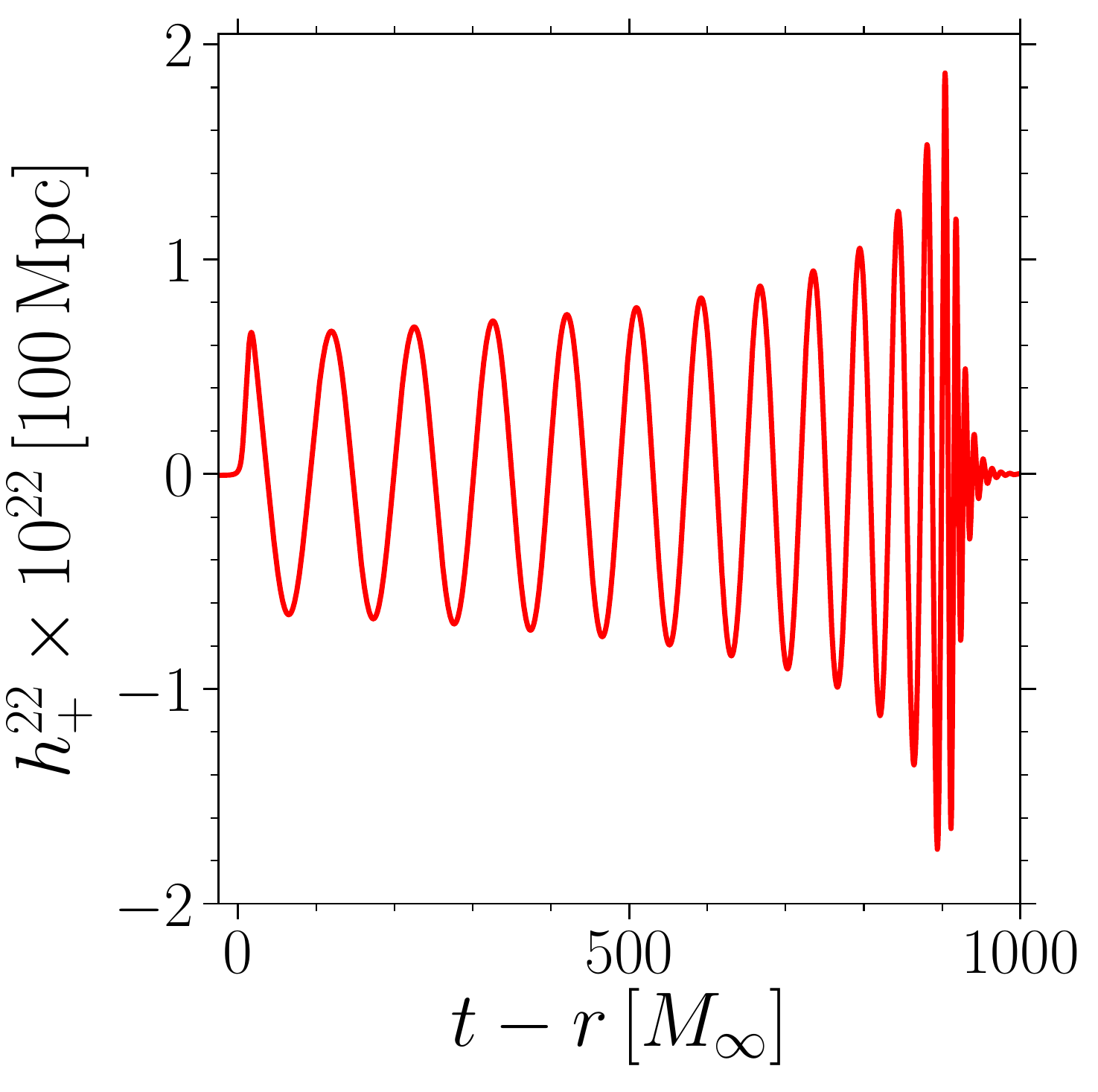}
  \caption{\textit{Left:} Orbital trajectories of a representative BBH
    configuration reproducing the properties of the GW150914 event; shown with
    the light-orange track is the orbit of the primary black hole, while the
    light-blue track refers to the secondary. \textit{Middle:} Evolution
    of the coordinate separation $r(t)$ of the GW150914 ID when
    considering only the quasi-equilibrium assumption (black solid line),
    the 3.5PN estimates for $\dot{a}$ and $\Omega$ (blue solid line), or
    after the fourth iteration (ECC4) of the eccentricity-reduction
    procedure (red solid line). The inset shows the time derivative of
    the coordinate separation, $\dot{r}(t)$, for the same
    datasets. \textit{Right:} gravitational-wave strain of the $\ell=m=2$
    multipole of the $+$
	polarization for the ECC4 dataset.}
    \label{fig:bbh:orbits}
\end{figure*}

\subsubsection{Representative mass ratio and mixed spins}
\label{sec:res:bbh:evo}

As a realistic test case to exercise the capabilities of the BBH ID
solver, we generate ID based on the GW150914 detection and thus assuming
that the mass ratio is $q = 0.8055$. The primary black hole is set to
have a dimensionless spin of $\chi_1 = 0.31$ and the secondary $\chi_2 =
-0.46$, while we fix the initial separation to $d_0 = 10\,M$; this setup
is very similar to the one used in Ref. \cite{wardell2016}. A summary of
the dynamics of this binary is offered in Fig. \ref{fig:bbh:orbits}, whose
different panels report, respectively, the orbital tracks (left panel),
the coordinate separation between the two black holes at different stages
of the eccentricity-reduction (middle panel), and the corresponding
gravitational-wave strain in the $\ell=m=2$ multipole of the $+$
polarization (right panel). Note that the left and right panels refer to
the configuration with the smallest eccentricity.

The spins of both black holes are perpendicular to the orbital plane and,
as a first step, we generate a corresponding dataset under the assumption
of quasi-equilibrium (QE) using Eq. \eqref{equ:virial}. As expected from
this raw ID, the actual evolutions reveal that the initial orbital
eccentricity is large, as can be can be seen from the black line in the
middle panel of Fig. \ref{fig:bbh:orbits}; in the same panel, the inset
provides a measure of the time derivative of the coordinate separation
$\dot{r}(t)$. Fortunately, this problem can be resolved rather
straightforwardly and already by simply utilising the 3.5PN estimates for
the expansion coefficient, $\dot{a}$ [\ie Eq. \eqref{equ:adot3PN}], and for
the orbital frequency, $\Omega$ [\ie Eq. \eqref{equ:Omega3PN}]. As shown
with the blue line in the middle panel of Fig. \ref{fig:bbh:orbits}, this
simple estimate already results in a greatly reduced orbital
eccentricity.

An additional reduction can be obtained after performing four iterations
of the eccentricity-reduction procedure described in Sec. \ref{sec:bbh:ecc}
and Appendix \ref{sec:ecc},
where we start from the 3.5PN ID until we obtain an orbital eccentricity
of the order of $10^{-4}$; we refer to this ID as ``ECC4'' hereafter. More
specifically, for each iteration of the eccentricity-reduction procedure,
the eccentricity is measured using the coordinate separation between the
centres of both horizons $r(t)$, and its time derivative $\dot{r}(t)$;
the two quantities are then fitted using the ansatzes
\eqref{equ:r-ecc-fit} and \eqref{equ:dr-ecc-fit}\footnote{Fitting $r$ and
$\dot{r}$ via \eqref{equ:r-ecc-fit} and \eqref{equ:dr-ecc-fit} obviously
yields two distinct estimates for the parameters associated to
Eqs. \eqref{equ:adot:corr} and \eqref{equ:omega:corr}. In practice we use
both of them to ensure reliable corrections, but, based on experience, we
utilise the corrections from $\dot{r}$ here.}. We note that both
quantities are measured during the first three orbital periods to ensure
a consistent measurement of the eccentricity, which, in turn, allow us to
obtain accurate corrections to the quantities $\Omega$ and $\dot{a}$ [\cf
  Eqs. \eqref{equ:adot:corr} and \eqref{equ:omega:corr}]. Experience has
shown that relying on a single orbit does not yield sufficiently accurate
estimates for corrections to $\Omega$ and $\dot{a}$, thus not yielding
a significant decrease in the eccentricity. In all cases, we are able
to obtain consistent measurements and corrections from $r(t)$ and
$\dot{r}(t)$ up to an eccentricity $\lesssim 10^{-3}$. For eccentricities
smaller than these and up to an eccentricity $\lesssim 10^{-4}$, we
obtain more reliable results using only the parameters fitted from
$\dot{r}(t)$, since the fitting parameters for $r(t)$ are unreliable due
to the eccentricity having a weak impact on the separation distance --
the oscillations are too small to fit -- when using the ansatz
\eqref{equ:r-ecc-fit}. Indeed, as remarked also by other authors
\cite{Pfeiffer:2007yz, Husa:2007rh, Buonanno2011, Kyutoku2014}, when
considering orbits with eccentricities $\lesssim 10^{-3}$, the correction
parameters are very sensitive to the fitting procedure used, to the
initial estimates for these parameters, and to the evolution window being
analysed.

\subsubsection{Impact of the ID resolution on the gravitational-wave phase}
\label{sec:res:bbh:error}

To further quantify the impact of the resolution with which the ID is
computed on the overall error budget as seen from an evolution
perspective, we run a series of nine simulations utilising the ECC4
initial dataset to determine the convergence of the gravitational phase
evolution up to merger. The latter is a good choice being a coordinate
independent quantity and the most important in waveform modelling for
template matching \cite{Hinder2013}.

This series of nine evolutions consists of a binary constructed with three
different \textit{ID resolutions}, \ie $\bar{N} = 24, 38$, and $42$, and
evolved with three different \textit{evolution resolutions}, \ie $\Delta
x_{\rm{LR}}/M = 0.024$, $\Delta x_{\rm{MR}}/M = 0.019$, and $\Delta
x_{\rm{HR}}/M = 0.015$. The latter correspond to a number of points
across the apparent horizon (AH) of about $n_{\rm{AH}} = 35$,
$n_{\rm{AH}} = 45$ and $n_{\rm{AH}} = 55$ respectively. For all cases
considered, the spacetime evolution utilises an 8th-order
finite-differencing scheme so as to minimise the error in the evolution
of the binaries.

\begin{table}[t]
  \def\arraystretch{1.75}
  \begin{centering}
\begin{tabular}{c|c|c|c||c|c}
 & $\left|\Delta\varphi\right|_{_{\rm LR}}$ &
  $\left|\Delta\varphi\right|_{_{\rm MR}}$ &
  $\left|\Delta\varphi\right|_{_{\rm HR}}$ & $M_{_{\rm ADM}}$
  $\left[{M}\right]$ & $J_{_{\rm ADM}}$ $\left[{M}^{2}\right]$\\
  \hline
  \hline
  $\bar{N}=24$ & $12.214$ & $1.888$ & $0.095$ & $0.9897$ & $0.9572$\\
  \hline
  $\bar{N}=38$ & $12.067$ & $1.771$ & $0.008$ & $0.9899$ & $0.9573$\\
  \hline
  $\bar{N}=42$ & $12.067$ & $1.770$ & $0.000$ & $0.9899$ & $0.9573$\\
\end{tabular}
\par\end{centering}
\caption{Gravitational-wave phase differences for the $\ell=m = 2$ strain
  mode of the $+$ polarisation as computed at merger when employing
  either different ID resolutions ($\bar{N} = 24, 38, 42$) or evolution
  resolutions (LR, MR, HR). Also reported are the corresponding values of
  the ADM mass $M_{_{\rm ADM}}$ and ADM angular momentum $J_{_{\rm ADM}}$
  as computed from the ID.}
\label{tab:bbh-conv-phi}
\end{table}

\begin{figure}[b]
  \includegraphics[width=.475\textwidth,keepaspectratio]{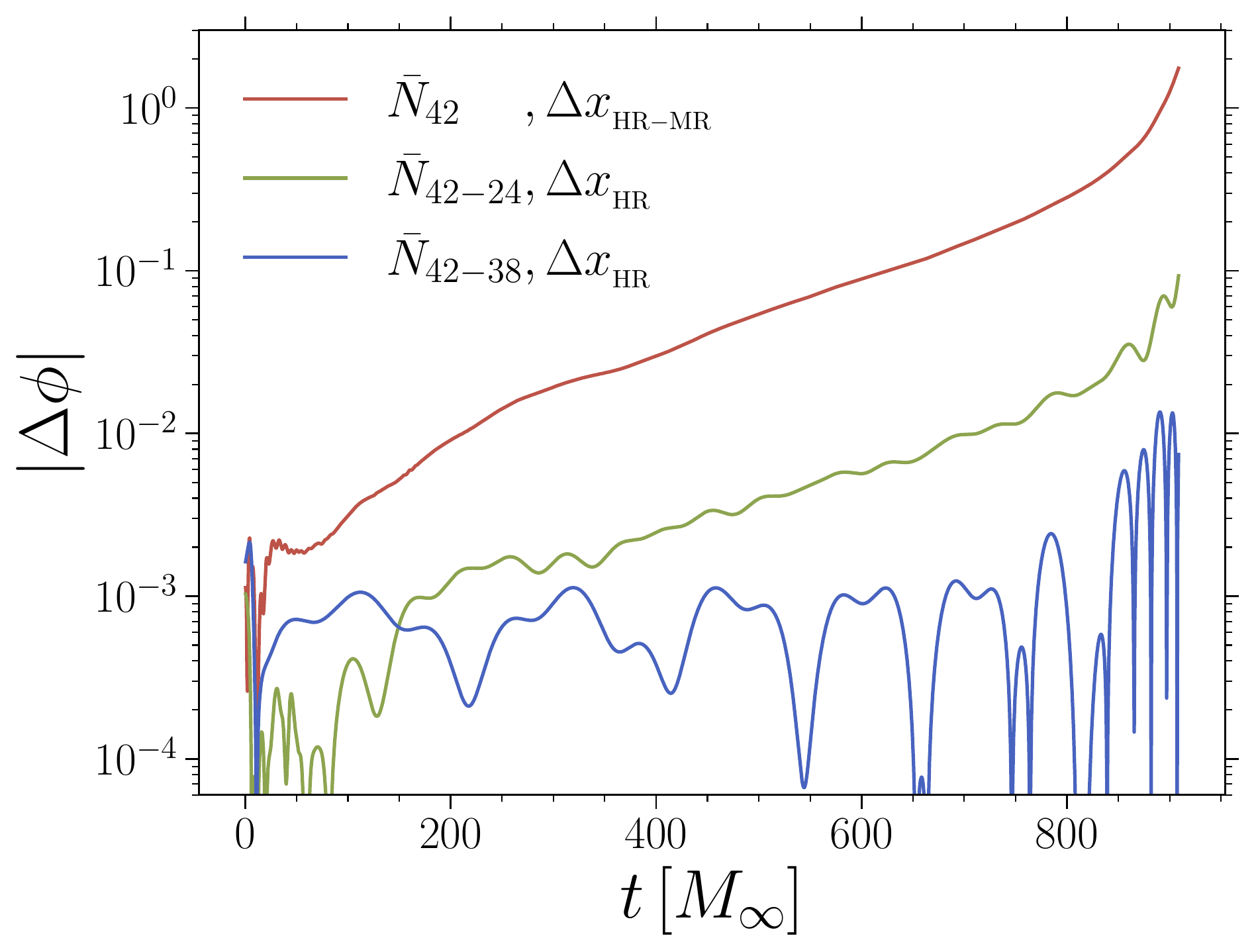}
  \caption{Evolution of the differences in the gravitational-wave phase
    computed from the $\ell=m = 2$ multipole of the $+$ polarization
    produced by BBH configurations representative of the GW150914
    event. Different lines contrast the difference when considering
    either different effective ID resolutions, \ie $\bar{N} = 24, 38,
    42$), or different evolution resolutions, \ie low resolution ($\Delta
    x_{\rm{LR}}$), medium resolution ($\Delta x_{\rm{MR}}$), and high
    resolution ($\Delta x_{\rm{HR}}$). Note that the contribution of the
    ID to the final error budget is always subdominant at the evolution
    resolutions employed here.}
    \label{fig:bbh:gws}
\end{figure}

In Tab. \ref{tab:bbh-conv-phi} we report the magnitude of the phase
differences at merger of the phases of the $\ell=m = 2$ mode
gravitational-wave strain. For each of the cases reported, $|\Delta
\phi|$ is computed as the difference between the gravitational-wave phase
at merger from evolutions at a given resolution (\ie LR, MR, HR) from ID
computed with a given set of collocation points (\ie $\bar{N}=24, 38,
42$) relative to the highest-resolution setup (\ie HR, $\bar{N}=42$). In
addition, and as a reference, Tab. \ref{tab:bbh-conv-phi} reports the 
various ADM quantities for each ID resolution.

Similarly, but only for a subset of three binaries in Tab.
\ref{tab:bbh-conv-phi}, we show in Fig. \ref{fig:bbh:gws} the full time
evolution of the phase differences. In particular, we concentrate on
evolutions capturing the differences of the ID datasets with $\bar{N} \in
\left\{24, 38 \right\}$ and evolved at the highest resolution HR. These
differences are indicated with blue and green lines in
Fig. \ref{fig:bbh:gws} and are meant to highlight the actual impact of
the ID resolution on the error budget of the simulation. In addition, we
report with a dark-red line the phase difference that develops when
comparing evolutions with ID computed at the highest resolution (\ie
$\bar{N}=42$) between the medium (MR) and high-resolution (HR) setups. By
contrast, this line is meant to highlight the actual impact of the
evolution resolution on the error budget.

As can be seen already from Fig. \ref{fig:bbh:gws} and fully deduced from
Tab. \ref{tab:bbh-conv-phi}, the total phase error at merger is
completely dominated by the evolution resolution, at least for the
resolutions considered here. There is only a very weak dependence on the
ID resolution, which converges away rapidly with increasing number of
collocation points. Stated differently, the ID error contribution is
subdominant already with $\bar{N}=24$ and becomes even less relevant as
the number of collocation points is increased. As customary in these
evolutions, the phase difference increases as the merger is approached
and evolution becomes increasingly nonlinear. However, even in the case
of the low-$\bar{N}$ ID, the phase difference is always below $\Delta
\phi \sim 0.1 \, \rm{rad}$. In contrast, the phase difference between the
two highest evolution resolutions is one magnitude larger, $\Delta \phi
\sim 1.0 \, \rm{rad}$, and is dominating over the whole inspiral. These
results clearly indicate that for vacuum solutions at the resolutions
considered here -- and for the ranges of mass ratios and spins explored
so far -- the ID resolution plays only a minor role for the total phase
error budget and rather low resolutions can be used as long as the
orbital frequency is fixed by PN estimates or iterative eccentricity
reduction. 

\subsection{Evolutions of neutron-star binaries}
\label{sec:res:bns}

We next present the results of the evolutions of BNS configurations whose
quasi-equilibrium initial configurations have been produced with the new
solver utilising the \texttt{Kadath} library. Also in this case, our
evolutions are here meant to be used mostly as representative test cases
and clear proofs of the capabilities of the new spectral-solver library
to produce astrophysically useful data, rather than providing new insight
into this process.

\begin{table}[t]
  \def\arraystretch{1.5}
  \begin{tabular}{|l|c|c|c|c|c|}
\hline
Reference & $\omega$ & $M_{b}$ & $M_{_{\rm QL}}$ & $\mathcal{S}_{_{\rm QL}}$ & $\mathcal{S}_{_{\rm QL}}/M^{2}$\\
 & $[M^{-1}_{\odot}]$ & $[M_{\odot}]$ & $[M_{\odot}]$ & $[M^{2}_{\odot}]$ & \\
\hline
\hline
Tichy+ 2019 \cite{Tichy2019}& $0.00000$ & $1.7745$ & $1.620$ & $-0.0007$ & $-0.0003$\\
\hline
this work & $0.00000$ & $1.7745$ & $1.620$ & $-0.0007$ & $-0.0003$\\
\hline
Tichy+ 2019 \cite{Tichy2019}& $0.01525$ & $1.7745$ & $1.626$ & $\phantom{-}0.8652$ & $\phantom{-}0.3217$\\
\hline
this work & $0.01525$ & $1.7745$ & $1.626$ & $\phantom{-}0.8631$ & $\phantom{-}0.3209$\\
\hline
\end{tabular}
  \caption{Comparison with the properties reported in
    Ref. \cite{Tichy2019} for either an equal-mass irrotational, or
    equal-mass spinning BNS configuration using a single polytrope with
    $K=123.6$ and $\Gamma = 2$. Listed are the quantities that can be
    compared directly: the fixed spin frequency parameter, $\omega$, the
    fixed baryonic mass, $M_{b}$, the quasi-local ADM mass, $M_{_{\rm
        QL}}$, the spin angular momentum, $\mathcal{S}_{_{\rm QL}}$, and
    the dimensionless spin normalized by $M=1.64\,M_{\odot}$. The
    agreement is very good and the small differences in the quasi-local
    measurements are mostly due to the different approaches to perform
    the measurements.}
\label{tab:spin-star}
\end{table}

\subsubsection{Spinning binary neutron stars: a comparison}
\label{sec:res:bns:comp}

As a first general test of a BNS system containing spinning companions,
we consider the equal-mass, equal-spin BNS model first presented in
Ref. \cite{Tichy2019}, which is based on a single polytrope with
$K=123.6$ and $\Gamma = 2$. A similar stellar model was considered also
in Ref. \cite{Tacik15}, but unfortunately no updated model was discussed
in the subsequent work Ref. \cite{Tacik16}. For this binary, the spin
parameter is fixed to $\omega = 0.1525$ [\cf Eq. \eqref{eq:enth:spin_2}],
together with a baryonic mass of $M_b = 1.7745\,M_{\odot}$, and a
coordinate separation of $d=47.2 \, \Msol$.

Table \ref{tab:spin-star} offers a comparison of the quasi-local
measurements for the mass and spin computed here with the corresponding
quantities reported in Ref. \cite{Tichy2019}. Note that while there is an
excellent agreement in the quasi-local mass computed by
\eqref{equ:adm-mass-ql}, there is a small deviation in the quasi-local
spin. We believe this difference is due to the method used in
Ref. \cite{Tichy2019} to compute the spin, which differs from the one
employed here and that follows the one in Ref. \cite{Tacik15}; the
differences are however minute and smaller than $0.3\%$.

To further assess the correctness of the implementation of the
spin-velocity field given by Eq. \eqref{eq:enth:spin_1} and the resulting
spin angular momenta, we created a sequence of equal-mass BNS models
based on a single polytrope with $K=123.6$ and $\Gamma = 2$. The
sequences are parameterized by the increasing spin parameter $\omega$ for
a fixed mass $M_{_{\rm ADM}} = 1.64\,M_{\odot}$, thus matching the models
given in \cite{Tacik15,Tacik16,Tichy2019}. Note that the baryonic mass
decreases for increasing spin at fixed $M_{_{\rm ADM}}$ due to the
growing contribution of the spin angular momentum to the gravitational
mass and, thus, has to be adjusted by matching it to single-star models
with the same $M_{_{\rm ADM}}$ and $\chi$.

\begin{figure}[t]
  \includegraphics[width=0.475\textwidth,keepaspectratio]{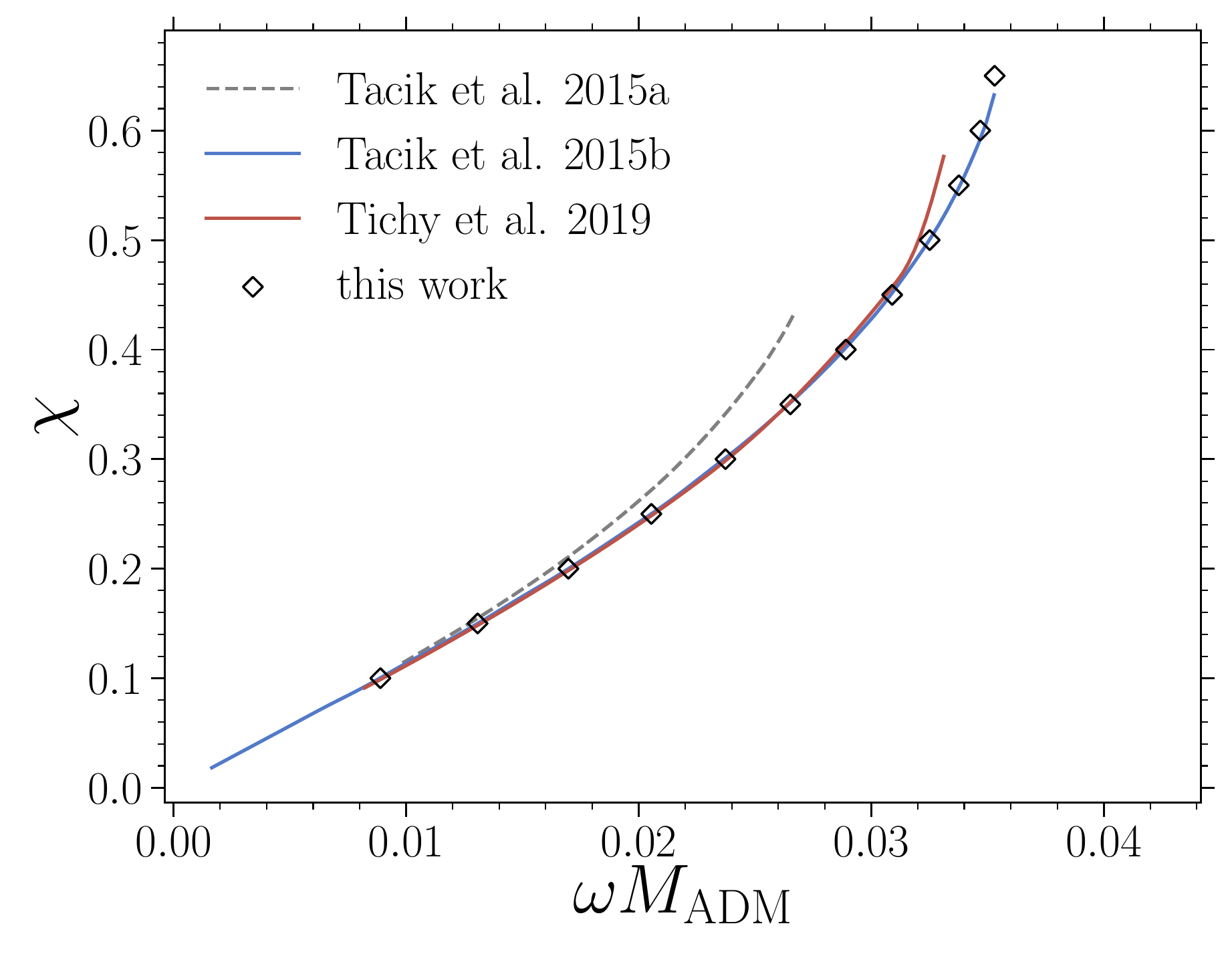}
  \caption{Dimensionless spin $\chi$ as function of the stellar spin
    frequency parameter $\omega$ for a sequence of BNS configurations
    using a single polytrope with $K=123.6$ and $\Gamma = 2$. The
    numerical data (open symbols) is compared with the interpolating
    functions reported in Refs. \cite{Tacik15,Tacik16,Tichy2019},
    indicating the very good agreement.}
    \label{fig:bns:chi-omega}
\end{figure}

The resulting dependency between the spin parameter $\omega$ and
dimensionless spin $\chi$ is shown in Fig. \ref{fig:bns:chi-omega} and
combined with a smoothly interpolated representation of the data given in
Ref. \cite{Tacik15,Tacik16,Tichy2019}; we note that the results
reported in Ref. \cite{Tacik15} (black dashed line in
Fig. \ref{fig:bns:chi-omega}) were generated with an incorrect
first-integral equation and has been corrected in Ref. \cite{Tacik16}
(blue solid line). It is evident from Fig. \ref{fig:bns:chi-omega} that
all three codes reproduce the same relation at low spin angular momenta
and that this is almost linear. However, for larger spin angular momenta
the relation becomes nonlinear with the spins increasing rapidly as
function of the frequency parameter. Note that for very high spins a
difference appears between the values computed here and those reported in
Ref. \cite{Tichy2019} (dark-red solid line). As discussed above, we
believe this discrepancy originates from different methods employed to
compute the quasi-local spin angular momentum; furthermore, since this
quantity is defined only approximately, the variations measured are not a
source of concern.

\subsubsection{Eccentricity reduction with unequal masses and spins}
\label{sec:res:bns:ecc}

As done for BBHs, we also employ an iterative eccentricity-reduction
procedure on our BNS ID that follows the same logic mentioned above and
presented in more detail in Appendix \ref{sec:ecc}. As it is natural to
expect, BNSs that are increasingly asymmetric in mass and spin exhibit an
increase in the initial eccentricity starting from the quasi-equilibrium
solution using the force-balance constraint equation
\eqref{eq:force-balance}. Especially in binaries with components with
large dimensionless spin, \ie $\chi \gtrsim 0.6$, the initial
eccentricity can be extremely large and becoming larger with increasing
spins and decreasing mass ratios.

As a general example of our eccentricity-reduction process involving
extreme spins, we generate a BNS configuration using the beta-equilibrium
slice of the finite-temperature TNTYST EOS \cite{Togashi2017} with
$M_\infty = 2.7\,M_{\odot}$, $q = 0.6875$, $\chi_1 = 0$, and $\chi_2 =
0.6$, where the highly spinning star is also the more massive
one.

Starting from the quasi-equilibrium solution, the eccentricity of the
orbit is progressively reduced via a total of four steps in which we use
the fitting ansatz \eqref{equ:dr-ecc-fit} for the time derivative of the
proper separation of both neutron stars. We remark that we employ
a Newtonian estimate for the barycentre of both stars to circumvent the
high-frequency noise in the location of the stellar centres that appears
when defining the stellar centres by a maximum density measurement
alone. The eccentricity reduction is performed using a lower resolution
ID with $\bar{N} = 29$ and a medium evolution resolution of $\Delta x =
0.2\, \Msol \, \approx 295 \, \rm{m}$. For the construction of the fourth
and final eccentricity-reduced dataset, the resolution is increased to
$\bar{N} = 38$. We note that further increasing/decreasing the resolution
of the ID between these two values of $\bar{N} = 29, 38$ at this stage of
the procedure has no substantial effect on the resulting evolution, as we
further discuss below (see Sec. \ref{sec:res:bns:error}).

\begin{figure}[t]
	\includegraphics[width=0.475\textwidth,keepaspectratio]{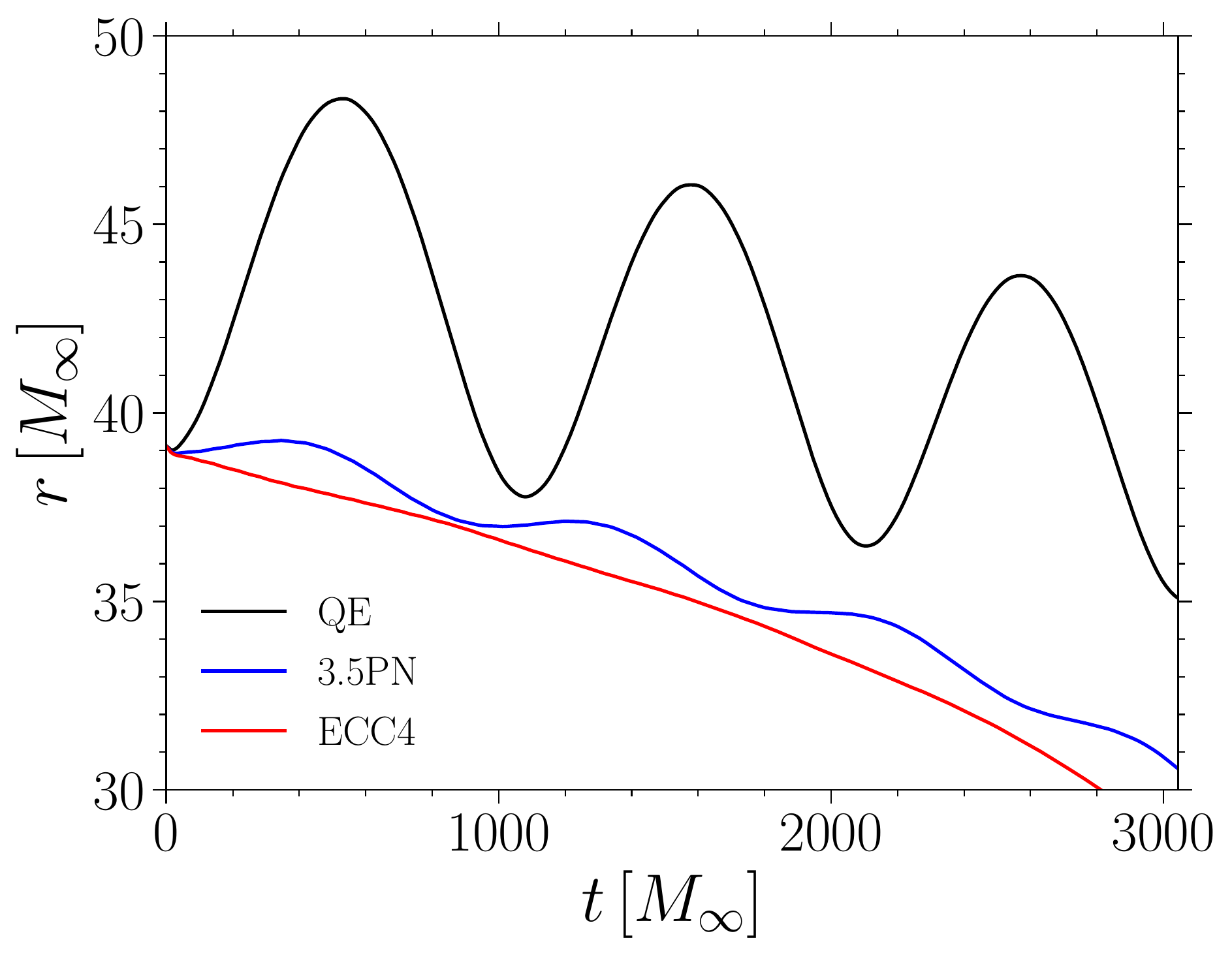}
	\caption{Representative example of the iterative eccentricity
          reduction for a rapidly spinning BNS system modelled with the
          TNTYST and with $M_\infty = 2.7$, $q = 0.6875$, $\chi_1
          = 0$ and $\chi_2 = 0.6$. Shown is the evolution of the proper
          separation between the two stars when using only the
          quasi-equilibrium ID (black line; QE), or when utilising the
          3.5PN estimates for $\Omega$ and $\dot{a}$ (blue line; 3.5PN),
          or when employing the ID from the final step of the
          eccentricity-reduction procedure (red line; ECC4). Note that
          the QE condition leads to enormous eccentricities for such a
          highly spinning binary.}
    \label{fig:bns:ecc-reduc-sep}
\end{figure}

In Fig. \ref{fig:bns:ecc-reduc-sep} we present the evolution of the
proper separation of the initial (black solid line) and final (red solid
line) datasets in the eccentricity reduction procedure\footnote{In
contrast to what happens with BBHs, whose proper distance is difficult to
calculate because of the inaccurate field values inside the AHs, the
actual proper distance can be calculated in the case of BNSs.}. In
addition, the same system is solved using fixed values of $\Omega$ and
$\dot{a}$ estimated from the 3.5PN expression given by
Eqs. \eqref{equ:adot3PN} and \eqref{equ:Omega3PN} (blue solid line),
which already provide a considerable reduction of the eccentricity. With
the final set of parameters we arrive at a residual eccentricity
$\lesssim 10^{-4}$, at which point the mentioned fitting procedure is no
longer reliable and further reduction becomes infeasible.

Figure \ref{fig:bns:ecc-reduc-sep} shows that the eccentricity-reduction
procedure performs very well even when starting with binary
configurations where the high spin of the more massive companion leads to
very large initial eccentricities.
At the same time, it is also apparent
that multiple iterations of the reduction can be skipped by simply
starting from the 3.5PN -- or higher-order PN estimates -- of the initial
orbital parameters. We thus recommend to apply these estimates in any
case instead of resorting to solutions based on the plain force-balance
equation \eqref{eq:force-balance} even when no further iterative
reduction is conducted. Indeed with very high spins as in this binary,
resorting to the 3.5PN expressions leads to eccentricities that are of
the same order as those encountered in standard irrotational
quasi-equilibrium configurations without eccentricity reduction.

\subsubsection{Impact of the ID resolution on the gravitational-wave phase}
\label{sec:res:bns:error}

In analogy with the results presented in Sec. \ref{sec:res:bbh:evo}, we
next investigate the impact of the ID resolution and of the evolution
resolution using the gravitational-wave phase as our reference
quantity. For this purpose, we conduct a series of simulations at varying
evolution resolutions, namely $\Delta x_{_{\rm LR}} = 0.25 \, \Msol
\approx 369 \, {\rm m}$, $\Delta x_{_{\rm MR}} = 0.2 \, \Msol \approx 295
\, {\rm m}$ and $\Delta x_{_{\rm HR}} = 0.145 \, \Msol \approx 215 \,
   {\rm m}$, in conjunction with three ID resolutions $\bar{N}_{_{\rm
       ILR}} = 29$, $\bar{N}_{_{\rm IMR}} = 38$ and $\bar{N}_{_{\rm IHR}}
   = 47$\footnote{In practice, we employ in each dimension an increment
   of four to the number of collocation points for the BNS ID in this
   case. Considering the exponential convergence of our spectral
   approach (see Fig. \ref{fig:bbh:convergence}), even such a small
   increase of collocation points leads to a nonlinear decrease of the
   truncation error.}. In particular, we concentrate on five combinations
   of these resolutions, considering first the two lower ID resolutions
   ILR and IMR and using them for the HR evolution resolution. Next, we
   compare and contrast the results to the highest resolution ID IHR,
   using it to perform evolutions at the three different evolution
   resolutions LR, MR and HR. As for the binary model, we resort to an
   equal-mass binary with individual baryonic masses $M_{b} = 1.4946 \,
   \Msol$ at an initial coordinate separation of $52.42\, {\rm km}$ using
   a tabulated version of the SLy EOS \cite{Douchin00}.

\begin{figure}[t]
  \includegraphics[width=0.475\textwidth,keepaspectratio]{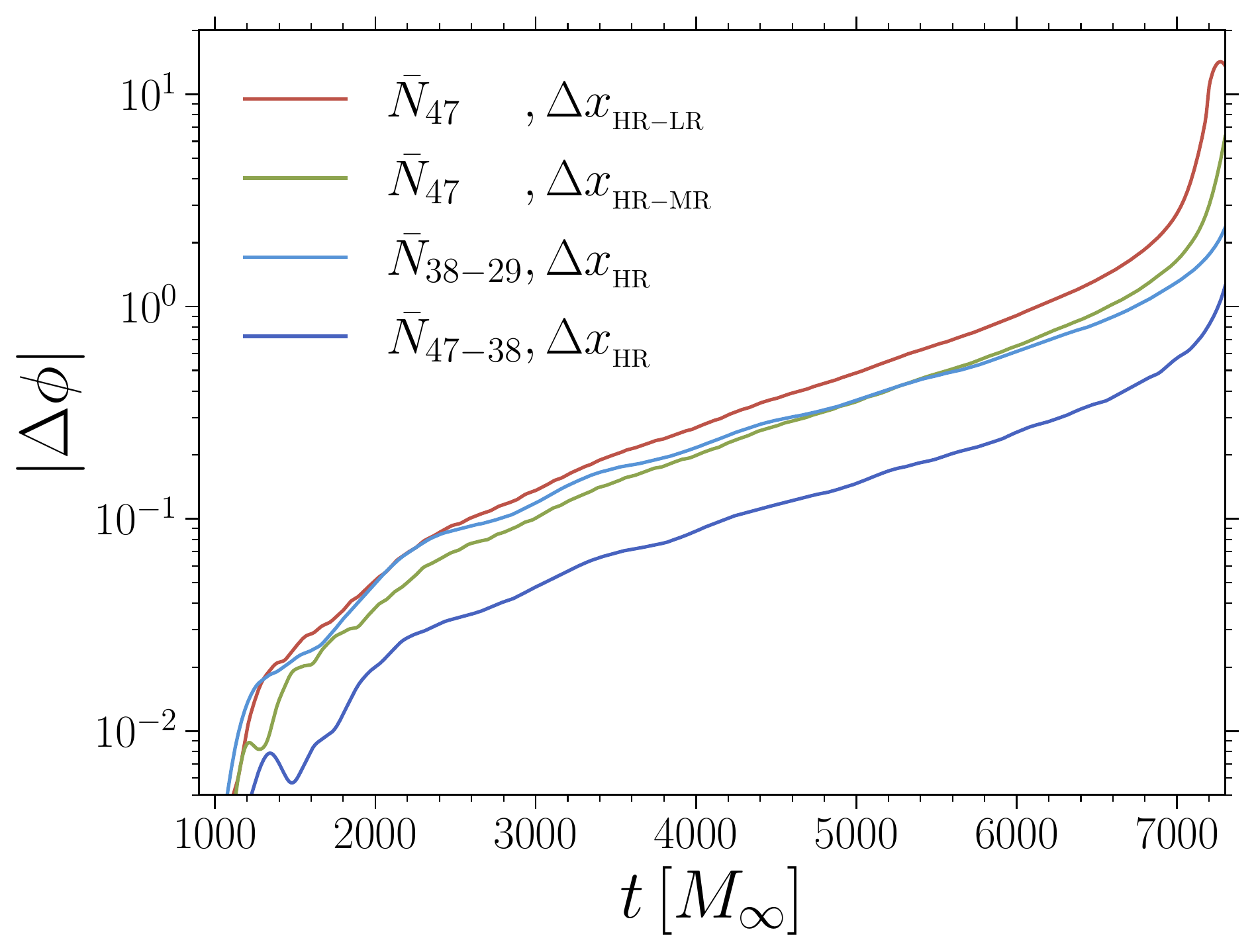}
  \caption{Same as in Fig. \ref{fig:bbh:gws}, but for an equal-mass
    irrotational BNS system modelled using the SLy EOS. Also in this
    case, the differences are computed either for different effective ID
    resolutions ($\bar{N} = 29, 38, 47$) or for different evolution
    resolutions (LR, MR, HR). Also in this case, the resolution evolution
    provides the largest contribution to the error budget at least for
    the resolutions considered here, although increasing the ID
    resolution can reduce the phase difference for HR evolutions.}
    \label{fig:bns:phi-diff}
\end{figure}

\begin{figure*}
  \includegraphics[width=0.70\textwidth,keepaspectratio]{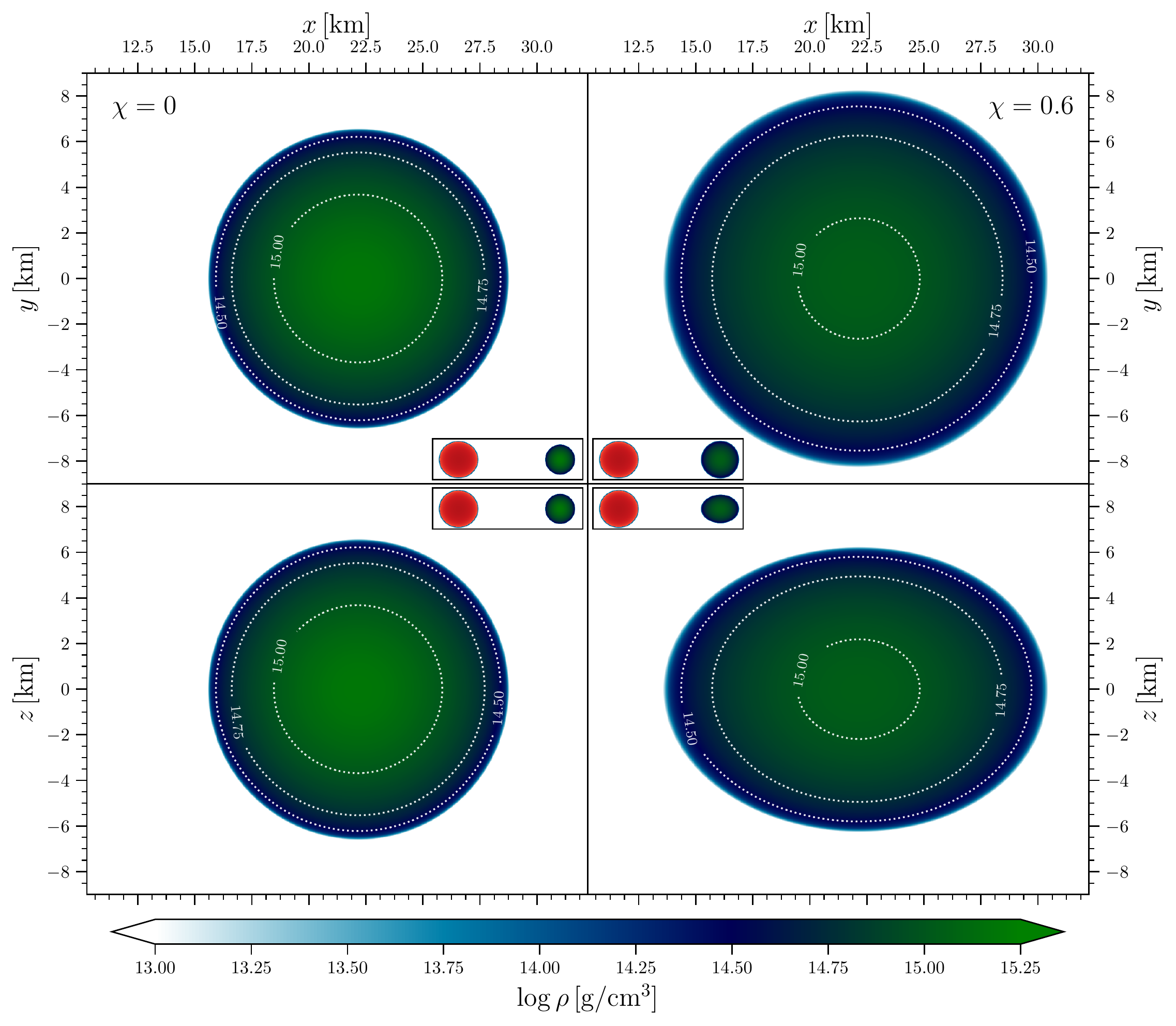}
  \caption{Two-dimensional cuts through the $(x,y)$ (top row) and $(x,z)$
    planes (bottom row) of two extreme BNS systems modelled with the
    TNTYST EOS and having a very small mass ratio ($q=0.455$
    corresponding to $M_1 = 2.2\, M_{\odot}, M_2 = 1.0\, M_{\odot}$). The
    left column refers to an irrotational binary ($\chi_1 = 0, \chi_2 =
    0$), while the right one to a very large spin asymmetry ($\chi_1 =
    0.6, \chi_2 = 0$); the latter is the most extreme BNS configuration
    considered here. The panels concentrate on the more massive
    component, but the insets offer views of the whole binaries, where
    the secondary is marked in red.}
    \label{fig:bns:2D}
\end{figure*}

We note that in order to remove effects of varying eccentricity at
different resolutions introduced by slightly changing orbital parameters
-- most notably, $\Omega$ -- we enforce a well controlled setup with
$\Omega$ and $\dot{a}$ fixed by Eqs. \eqref{equ:Omega3PN} and
\eqref{equ:adot3PN}, respectively. An alternative route would be to
perform a full eccentricity reduction of the orbit to fix both
parameters.

As discussed in Sec. \ref{sec:res:bbh:error}, for each simulation we
compute the phase evolution of the $\ell=m=2$ mode gravitational-wave
strain and present in Fig. \ref{fig:bns:phi-diff} the resulting phase
errors. We note that -- in contrast with what is done for BBHs, where this was
not necessary -- we exclude the initial phase of the evolution, as the
binaries settle down after the junk is radiated away and we align the
waveforms at $1000 \, \Msol$. When considering the variations in the
phase evolution reported in Fig. \ref{fig:bns:phi-diff}, a few
considerations can be made. First, the largest differences in $\Delta
\phi$ are measured when considering differences in the evolution
resolution (dark-red and green solid lines), with the difference when
considering the HR and LR resolutions (dark-red solid line),being larger
than when considering the HR and MR resolutions (green solid line).
In other words, and as already commented above, the resolution evolution
provides the largest contribution to the error budget and having large ID
resolution does not provide a more accurate phase evolution for the
evolution resolutions considered here. Second, the smallest values of
$\Delta \phi$ are obtained when considering the highest evolution
resolution and the two largest ID resolutions (dark-blue solid
line). Third, using a low ID resolution, \ie $\bar{N} = 29$, but high
resolution evolution is already sufficient to obtain an overall
difference that is comparable with that obtained with much higher ID
resolution, \ie $\bar{N} = 29$, but coarser evolution resolution
(light-blue solid line). Finally, note that all the phase differences
have roughly the same growth rate, once again indicating that the largest
source of error is not the calculation of the ID, but rather the
resolution employed in the evolution and, of course, the order of the
numerical method employed in the evolution part\footnote{We have here
employed a 4th-order spatial finite-difference scheme for the BNS
spacetime evolution. This is appropriate, since the effective convergence
order of the hydrodynamics solver, which is $<3$, will determine the
accuracy of the results \citep{Most2019b}.}.

\begin{figure*}[t]
  \includegraphics[width=0.33\textwidth,keepaspectratio]{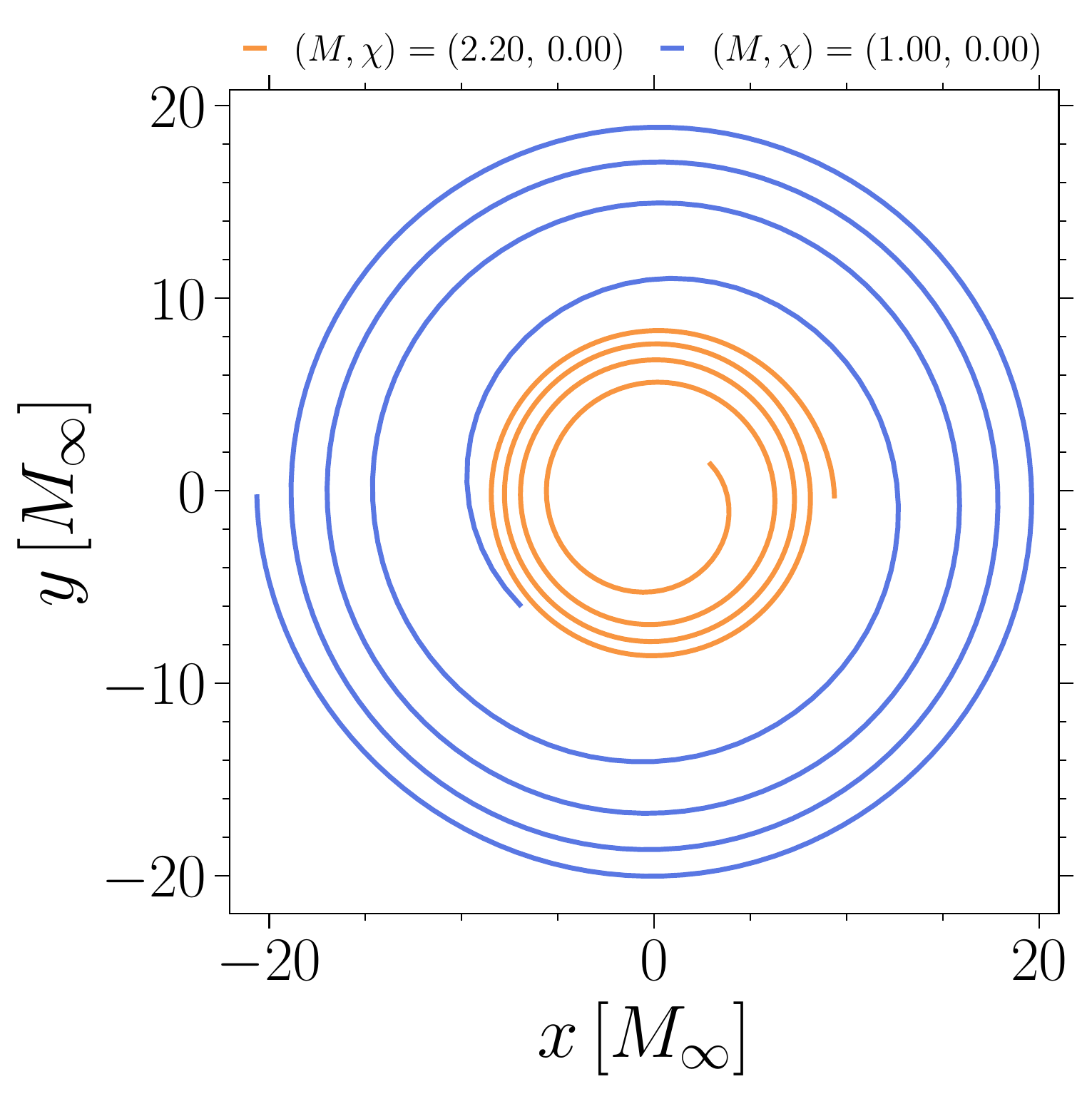}
  \includegraphics[width=0.33\textwidth,height=0.313\textwidth]{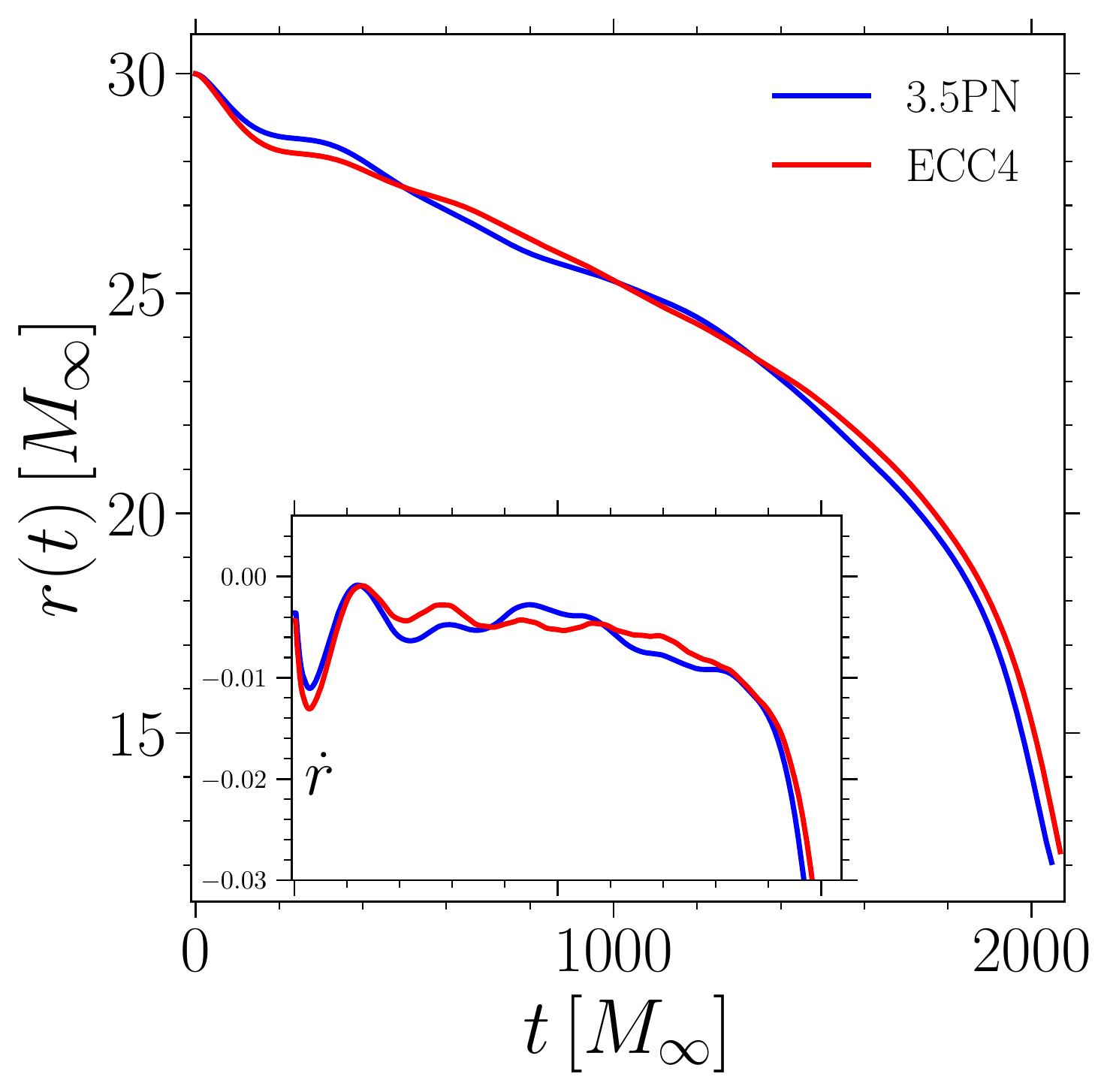}
  \includegraphics[width=0.33\textwidth,keepaspectratio]{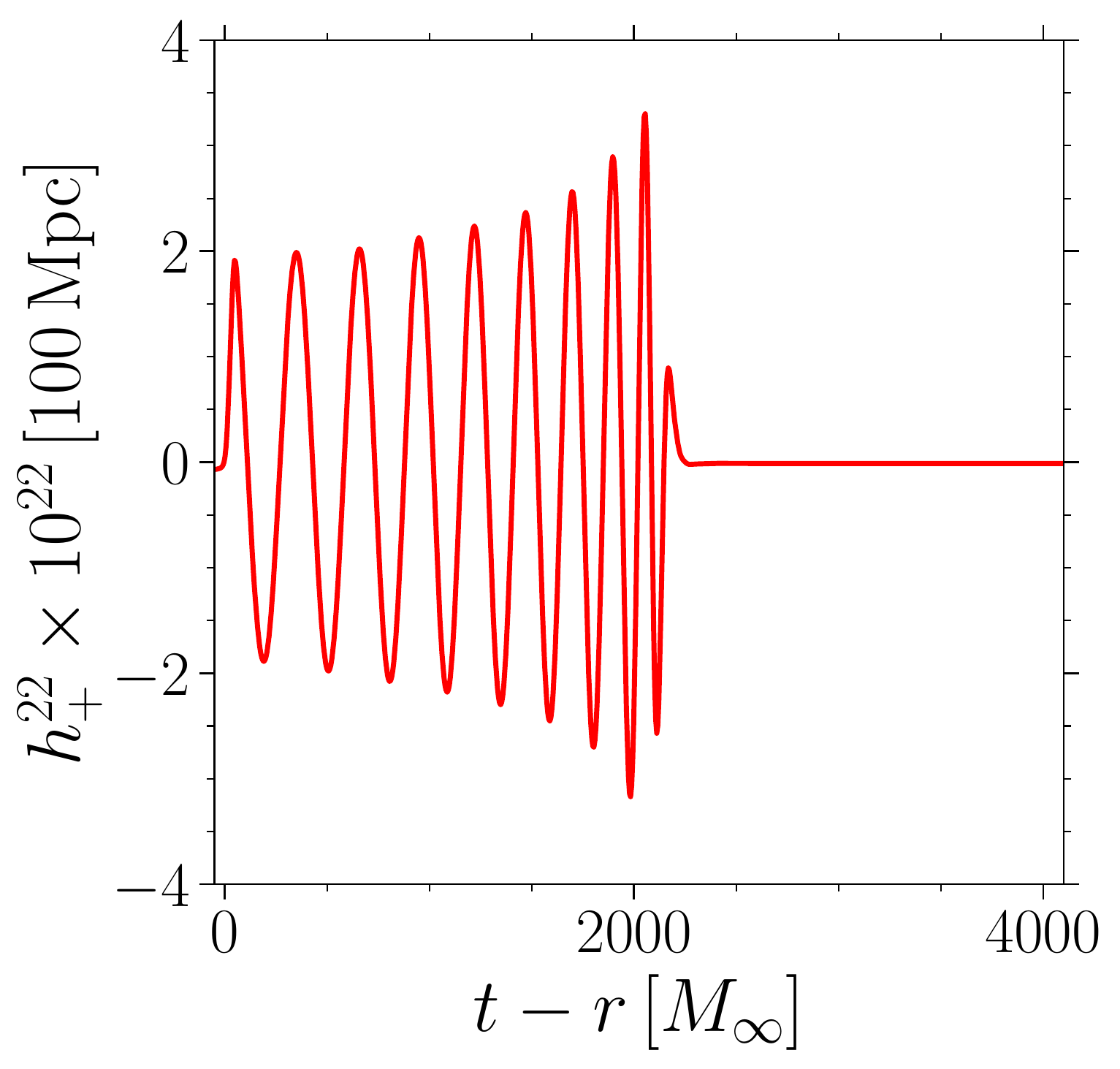}
  \includegraphics[width=0.33\textwidth,keepaspectratio]{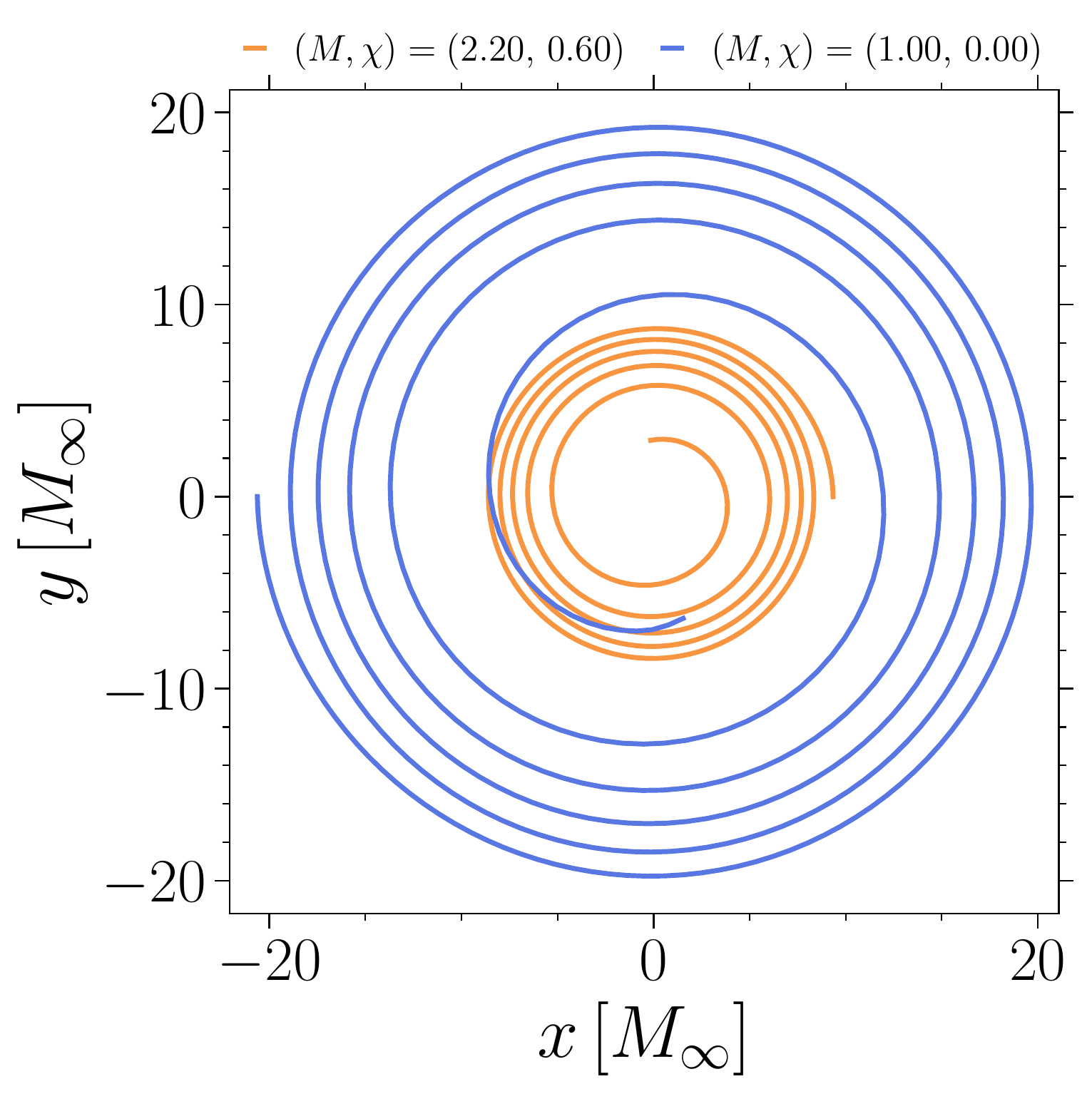}
  \includegraphics[width=0.33\textwidth,height=0.313\textwidth]{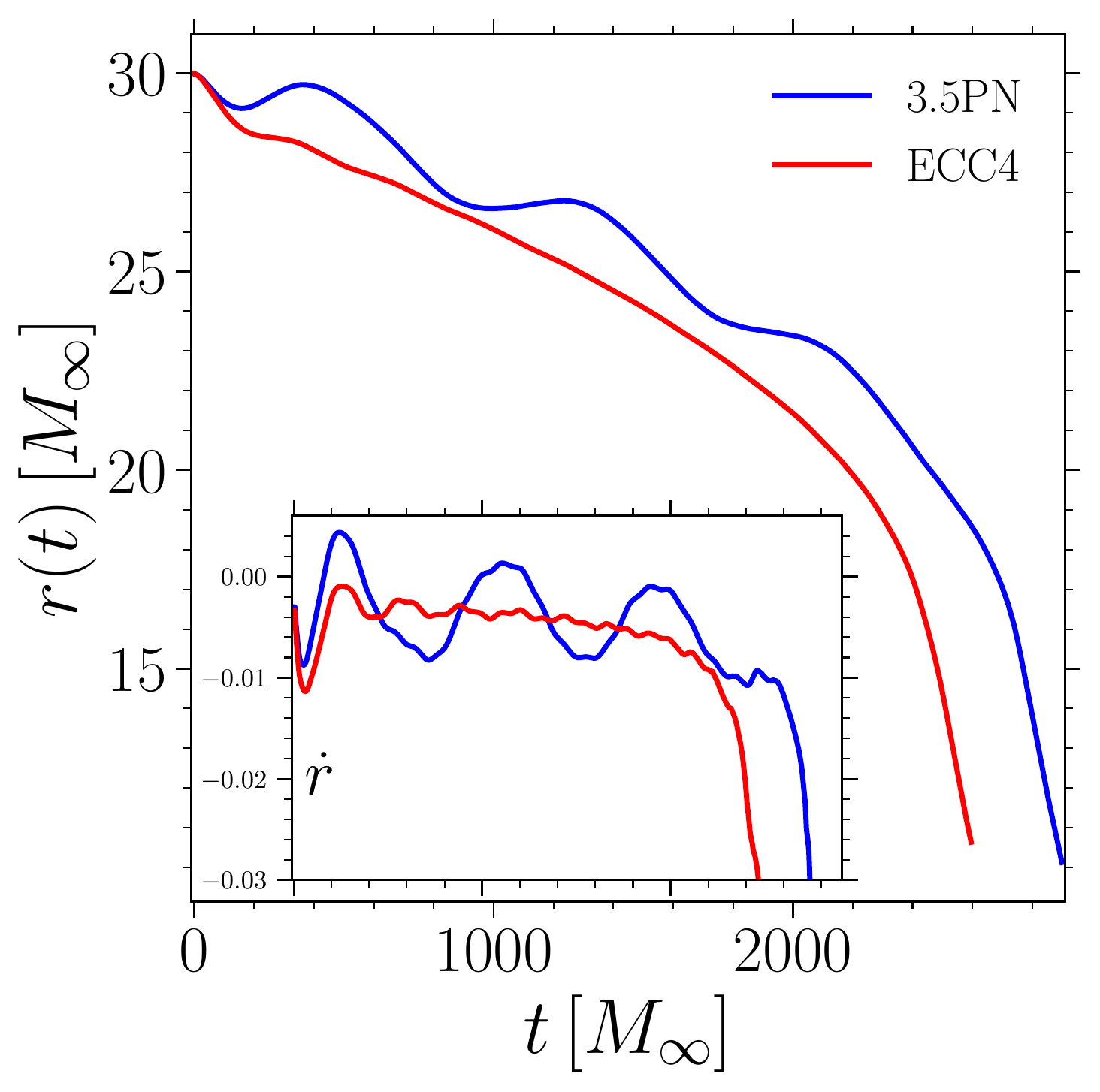}
  \includegraphics[width=0.33\textwidth,keepaspectratio]{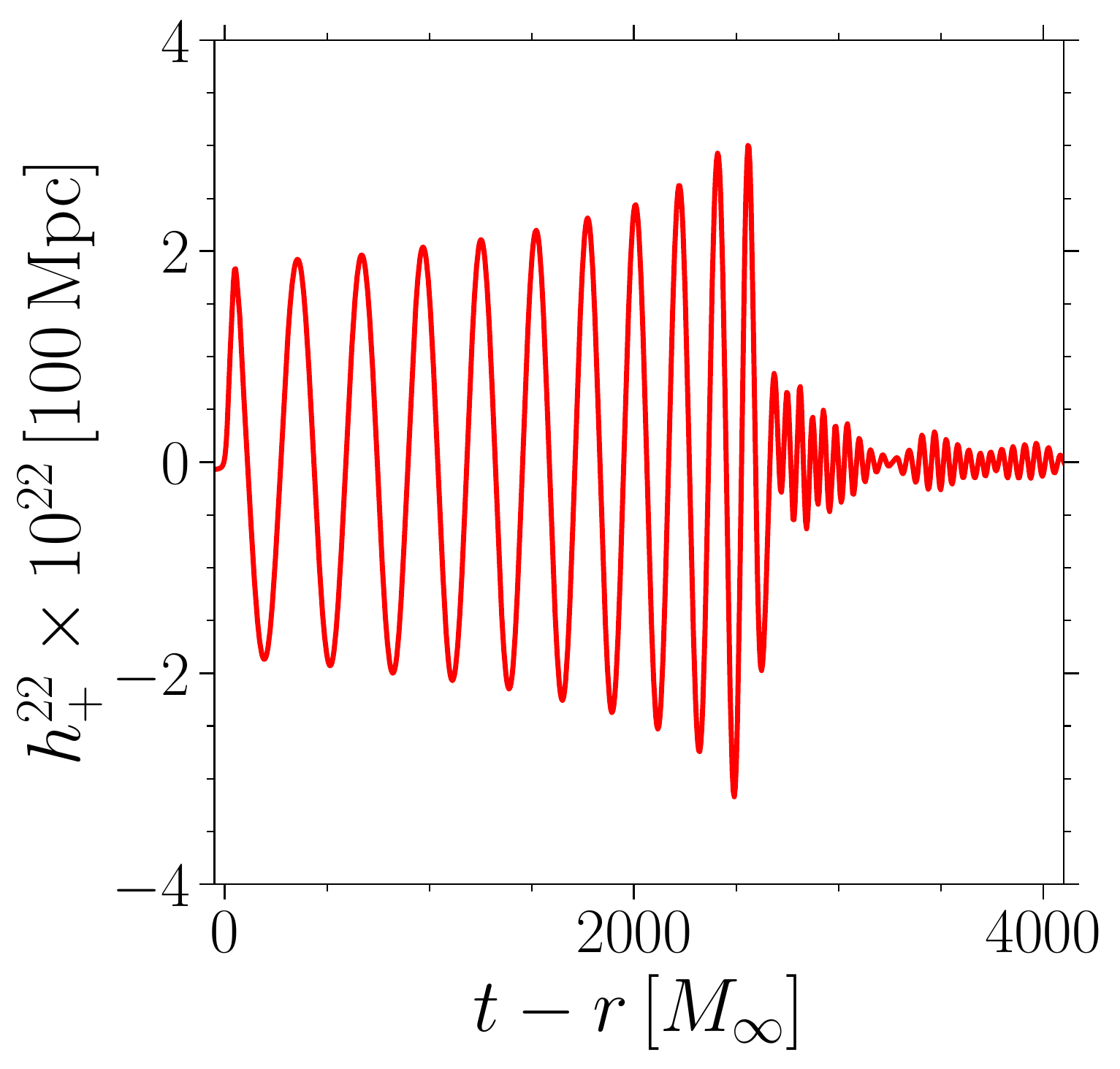}
  \caption{Same as in Fig. \ref{fig:bbh:orbits} when referring to extreme
    BNS configurations modelled with the TNTYST EOS (see also
    Fig. \ref{fig:bns:2D}). The top row reports the orbital trajectories,
    the evolution of the proper separation after different eccentricity
    reductions, and the gravitational-wave strain for a irrotational BNS
    with mass ratio $q=0.455$. The bottom row reports the same quantities
    for a BNS with the same mass ratio but extreme spin asymmetry,
    $\chi_1 = 0, \chi_2 = 0.6$. Note that the large angular momentum of
    the spinning binary leads to more orbits and to a metastable merged
    object rather than to a black hole.}
    \label{fig:bns:orbits}
\end{figure*}

From there on, we follow the phase difference between the evolution as
well as the ID resolutions compared to the highest resolution simulation.
Both, the low evolution and ID resolution configurations are dominating
the phase error in the early inspiral, while the higher resolution ID
starts off with a significantly lower phase error. The slope of the
growth of both contributions to the error over time is slightly differing
and the evolution error is exceeding the accumulated errors from the low
resolution ID towards merger, \ie $\Delta \phi > 1$. While the error
using very low resolution ID is still comparable, using higher resolution
ID leads to significantly smaller phase errors at merger when compared to
the pure evolution error, being $\Delta \phi \approx 1$.

Overall, the result of these numerous simulations indicate that the error
on the phase evolution introduced by the ID obtained with $\bar{N} \ge
38$ should be smaller than the typical error introduced by the evolution,
especially for long inspirals. At the same time, increasing the ID
resolution for evolutions at very high resolutions can improve the
accuracy of the waveforms and yield a phase-evolution error that is
$\Delta \phi \approx 1$. While a more thorough investigation covering
larger portions of the parameter space is necessary for a precise picture
of the error budget, it is already clear that that ID involving source
terms like a perfect fluid demands higher evolution resolutions in
general (\cf Sec. \ref{sec:res:bbh:error}).
\vfill
\subsubsection{Extreme mass ratios and spins}
\label{sec:bns:extreme}

As a final capability test of the new BNS ID spectral-solver, we consider
two configurations that are at the edges of the physically plausible
space of parameters, thus generating two particularly extreme
configurations. More specifically, we consider binaries built with the
TNTYST tabulated EOS and create a first binary configuration at a
separation of $30\,\Msol$, with a mass ratio of $q=0.455$ and individual
masses $M_{1} = 2.2 \, \Msol, M_{2} =1.0\, \Msol$, so that the total mass
of the binary is $M_{\infty} = 3.2 \, \Msol$\footnote{We recall that the
TNTYST EOS has a maximum TOV mass of $M_{_{\rm TOV}} = 2.23\,M_{\odot}$,
so that the more massive component of the binary is very close to this
limit in the irrotational case.}. To the best of our knowledge, this is
represents the BNS configuration with the smallest mass ratio ever
computed.

\begin{figure*}[htb!]
  \includegraphics[height=.330\textwidth,keepaspectratio]{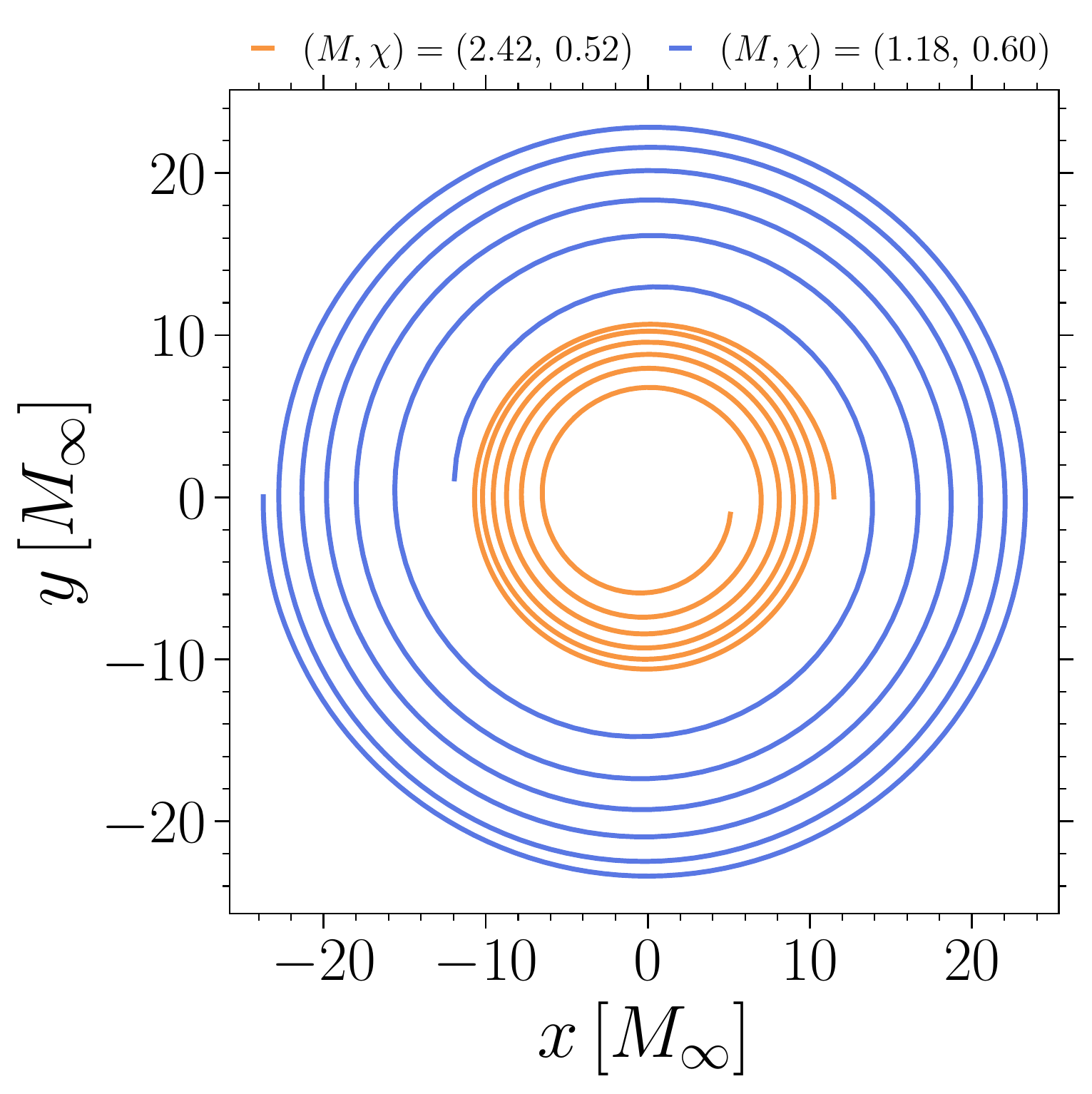}
  \includegraphics[height=.315\textwidth,keepaspectratio]{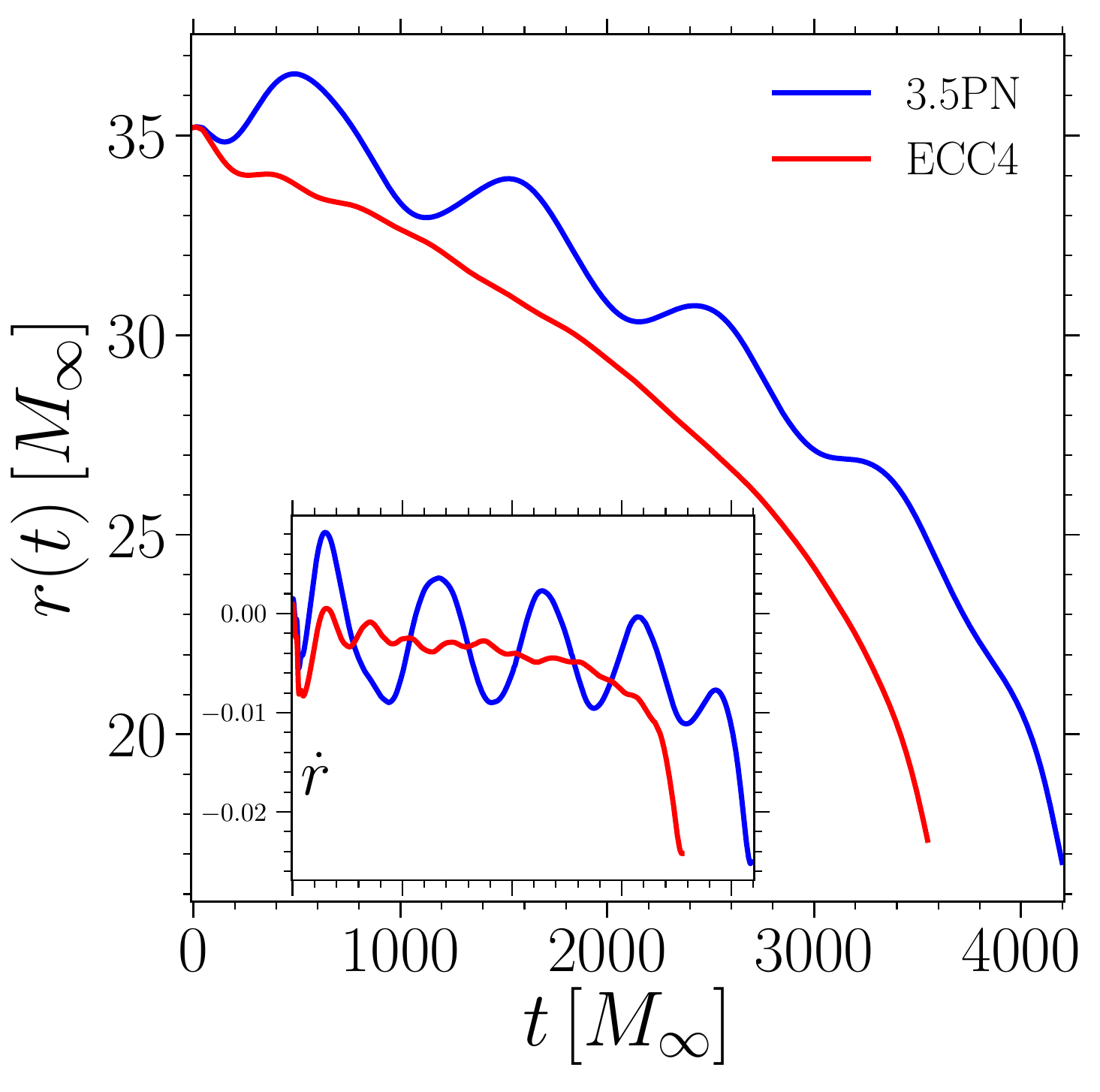}
  \includegraphics[height=.315\textwidth,keepaspectratio]{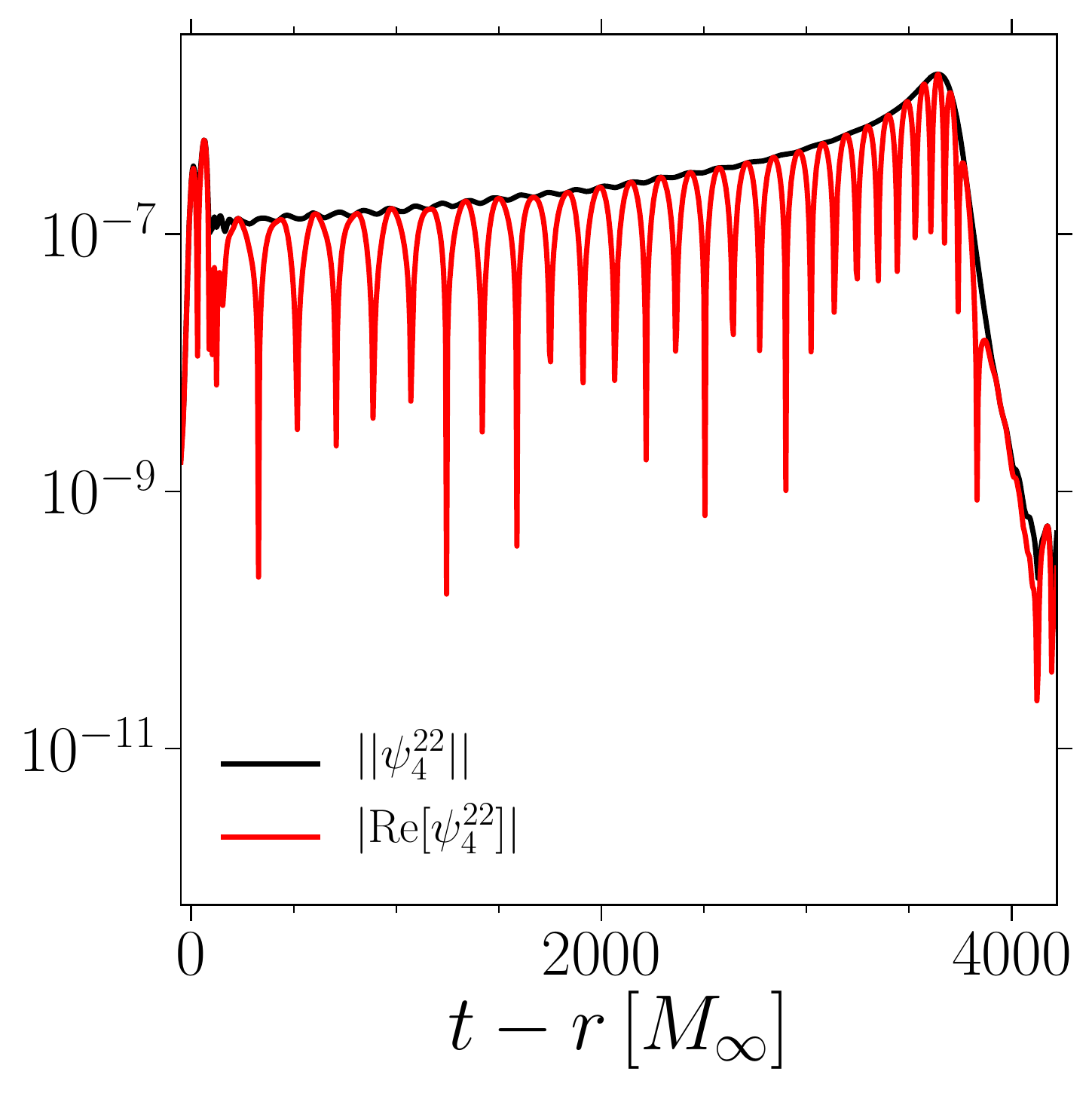}
  \caption{Same as in Figs. \ref{fig:bbh:orbits} and \ref{fig:bns:orbits}
    but for a BHNS configuration. Note that the right panel reports the
    $\ell=m=2$ multipole the $\psi_4$ Weyl scalar [(both the real part
      (red solid line) and its norm (black solid line)] in order to
    highlight the very short ringdown that would not be visible in the
    gravitational-wave strain.}
    \label{fig:bhns:gworbit}
\end{figure*}

Given these masses, we create one BNS configuration with both stars being
irrotational, \ie $\chi_1=\chi_2=0$, and a corresponding configuration
where the more massive companion is spinning extremely rapidly and the
less massive component is nonspinning, \ie $\chi_1=0.6, \chi_2=0.0$. This
second BNS configuration could be seen as a realisation of a recycled
binary pulsar in which one star gained a significant amount of matter and
angular momentum through an exceptional accretion phase. It is important
to remark that a binary configuration with unequal mass and unequal
spins, as the one considered here, is more challenging to compute than
when the masses are the same or when the spins are the same or, in
general, of smaller magnitude.

Interestingly, despite this being a rather extreme configuration, the
solver was able to generate this ID accurately and without any particular
fixes or changes to the equations discussed in
Sec. \ref{sec:init.data.matter}. Indeed, already at a very low resolution
of $\bar{N} = 19$ we were able to generate a fully converged solution,
which was successively scaled up directly to and resolved at a resolution
of $\bar{N} = 47$. Finally, before performing the evolution of these BNS
configurations, we employed the iterative eccentricity reduction
procedure using as reference the low-resolution dataset and thus reaching
an orbital eccentricity of $\lesssim 10^{-4}$.

Figure \ref{fig:bns:2D} provides a direct measure of the properties of
the two configurations by offering a cut through the $(x,y)$ and $(x,z)$
plane of the rest-mass density of the more massive star. The figure is
organised in four panels, with the left column referring to the
irrotational binary [$(x,y)$ plane on the top row and $(x,z)$ plane on
the bottom row], while the right column reports the spinning
binary. Also, we employ contour lines around the highest densities 
reached to help locate the most massive parts of the two stars
and include small insets that are representations of the two
binaries (the less massive companion is marked in red). 
As expected, the rapidly spinning
star is strongly flattened, extending further out along on the equatorial
plane and having a smaller extent along the $z$-axis. Furthermore,
because of this distortion, the nuclear region of rest-mass density $\rho
> 10^{15} \, \rm{g/cm^3}$ is smaller as in the irrotational model,
despite having the same $M_{_{\rm ADM}}$ at infinity.

We evolve both systems and present their trajectories in
Fig. \ref{fig:bns:orbits} following the same convention for the
quantities reported in Fig. \ref{fig:bbh:orbits}. Here, however, the top
row refers to an irrotational binary, while the bottom row reports the same
quantities for a BNS with extreme spin asymmetry, $\chi_1 = 0.6, \chi_2 =
0$. It is evident that the system with the highly spinning massive
companion takes longer to merge. With the given fixed initial separation
the difference amounts to approximately one orbit, which is due to the
larger total angular momentum to be radiated away prior to the merger
\cite{Kastaun2013, Dietrich2017c, Ruiz2019, Most2019}. Finally, in the
right panel of Fig. \ref{fig:bns:orbits} we report the corresponding
gravitational-waves strains in the $\ell=m = 2$ mode and $+$
polarization. An important consideration to make here is that the binary
system having the rapidly spinning companion is not collapsing promptly
(red solid line in the right panel of Fig. \ref{fig:bns:orbits}), in
contrast to what happens for the irrotational binary (black solid line),
whose merged object collapses right after merger. This behaviour clearly
suggests that spins can have a potentially important impact in
determining the threshold mass to prompt collapse \cite{Bauswein2013,
  Koeppel2019} and thus need to be properly modeled to obtain accurate
estimates of such masses over the entire physically relevant part of the
space of parameters.

\subsection{Evolutions of black-hole--neutron-star binaries}
\label{sec:res:bhns}

As a final application of our new solvers and as an additional example of
its flexibility, we consider the generation of ID representing a BHNS
system. More specifically, we have considered a BHNS binary with a mass
ratio of $q=0.485$ and a separation of $35.2\, \Msol$ together with
aligned spins of $\chi_{_{\rm BH}} = 0.52$ and $\chi_{_{\rm NS}} = 0.60$
(see also \cite{Most2020e}) utilising the TNTYST EOS \cite{Togashi2017}
to model the nuclear matter. The initial orbital frequency $\Omega$ and
the radial inward velocity of the orbit are fixed by the 3.5PN estimates
given by Eqs.\eqref{equ:adot3PN} and \eqref{equ:Omega3PN}, using
$M_{_{\rm ADM}}$ of the neutron star in isolation and $M_{_{\rm CH}}$ of
the black hole as measured on the horizon. Although the two spins are
neglected in these first estimates, they yield sufficiently reasonable
initial guess with which to begin the eccentricity reduction
procedure. Indeed, after performing four iterative steps, the final
eccentricity of our binary is $\lesssim 10^{-3}$, where the resulting
corrections for the first three steps of the iterative procedure were
obtained using both the coordinate separation $r(t)$ and its derivative
$\dot{r}(t)$. We note, however, that for eccentricities below $\sim 5
\times 10^{-3}$, the corrections based on $r(t)$ lead to an increasingly
eccentric orbit, so that the final (ECC4) ID dataset was obtained using
estimates based on $\dot{r}(t)$ only.

In analogy with Figs. \ref{fig:bbh:orbits} and \ref{fig:bns:orbits}, we
report in Fig. \ref{fig:bhns:gworbit} the orbital trajectories, the
evolution of the proper separation after different eccentricity
reductions, and the gravitational-wave strain for the BHNS ID with mass
ratio $q=0.485$. Note that the system undergoes six orbits in total and
exhibits a very low residual eccentricity throughout the inspiral (middle
panel). Furthermore, the center of mass that can be deduced already from
the orbital tracks stays at the origin of the simulation domain,
indicating a successful removal of the total residual linear momentum of
the spacetime. Finally, note also that the gravitational-wave signal has
a sharp cutoff after merger due to the disruption of the neutron star
(this was remarked also in Ref. \cite{Foucart:2020xkt}). To highlight
this behaviour and to reveal the ringdown, we do not report in the right
panel of Fig. \ref{fig:bhns:gworbit} the gravitational-wave strain in the
$\ell=m=2$ multipole, but the corresponding multipole of the $\psi_4$
Weyl scalar [(both the real part (red solid line) and its norm (black
  solid line)]. It is clear that in this case the ringdown is very
visible even if restricted to a couple of oscillations.

As a final remark we note that while our exploration of the space of
parameters with the new solver is certainly very limited and aimed
mostly at obtaining some reference solution, the calculation of BHNS ID
has been successful for all of the cases we have explored and that have
been restricted to black-hole spins $\chi \lesssim 0.75$, for which the
conformal flatness is still a reasonable assumption. Moving to
higher-spin black hole may require additional tuning since it is well
known that the conformally flat background metric is not able to reliably
reproduce highly spinning black-hole solutions (see
\cite{Lovelace2008c}).

\section{Discussion}
\label{sec:discussion}

A considerable effort has been dedicated in recent years to the
construction of accurate and realistic initial data representing generic
configurations of compact-object binaries in quasi-equilibrium. These
configurations -- which can either be of two black holes, of two neutron
stars, or of a black hole and a neutron star -- have then been employed
for successful evolutions, starting from the early inspiral and well past
merger. All of these simulations have enriched our understanding of
merging binaries and helped in the interpretation of the signal from
gravitational-wave detectors such as LIGO and Virgo.

While there are laudable examples of publicly available codes generating
this type of initial data, these codes often provide only a limited
capability in terms of mass ratios and spins of the components in the
binary. In particular, there is at present no open-source code including
the treatment of spinning neutron stars and an efficient procedure for
the reduction of the initial eccentricity. In addition, there also exists
a portion of the space of parameters -- namely, the one considering the
combination of extreme mass ratio and extreme and possibly differing
spins for systems of binary neutron stars -- that has, to date, not been
explored in the context of constraint-satisfying initial data.

The work presented here aimed at filling this gap by providing an
open-source collection of elliptic solvers that are capable of exploring
a major part of the space of parameters relative to binary black holes
(BBHs), binary neutron stars (BNSs), and mixed binaries of black holes
and neutron stars (BHNSs). The starting point of our development has been
the \texttt{Kadath} library, which is a highly parallelised spectral
solver designed for numerical-relativity
applications\cite{Grandclement09}. In addition, it is equipped with a
layer of abstraction that allows equations to be inserted in a
\LaTeX-like format.

The set of elliptic equations employed for the calculation of the ID is
well-known and has been presented in a number of related works. More
specifically, we employ the extended conformal thin-sandwich method
(XTCS), where the presence of a black hole is modeled by the usual
excision approach using particular inner boundary conditions on the
horizons, while the presence of a neutron star is modeled by either pure
irrotational or with an additional rotational velocity contribution. By
supporting both analytic EOSs, \eg single polytropes and piece-wise
polytropes, but also tabulated EOSs at zero or finite temperature, the
new infrastructure is particularly geared towards allowing for the
construction of BHNS and BNS binaries. For the latter, we showed that the
new spectral solvers are able to reach the most extreme corners in the
physically plausible space of parameters, including extreme mass ratios
and spin angular momenta, the most extreme computed to date. For a first
application of such extreme configurations with stellar companions close
to their maximum mass $M_{_{\rm TOV}}$ see \cite{Most2020e}. In this work
we went even further and presented for the temperature-dependent TNTYST
two BNS systems with an extreme mass asymmetry of $q = 0.455$, a primary
component with mass very close to the maximum mass \ie $M_1 / M_{_{\rm
    TOV}} > 0.98$. These binaries are either irrotational or with large
spin asymmetry, where the primary is very rapidly rotating with $\chi_1 =
0.6$. To the best of our knowledge, this is the most extreme BNS
configuration computed to date.

As illustrated in terms of a systematic series of examples, the new
spectral-solvers are able to construct quasi-equilibrium and
eccentricity-reduced ID for BBHs, BNSs, and BHNSs, achieving spectral
convergence in all cases. Furthermore, to assess the correctness of the
newly constructed binary configurations, we have carried out evolutions
of these systems from the inspiral to after the merger, obtaining in all
cases a behaviour consistent with the expectations and previous results.
An important aspect of these evolutions has been represented by the
construction of ID that has only a minimal amount of initial
eccentricity. The latter can be particularly large in the case of BNSs
with small mass ratios and containing rapidly spinning companions, but is
suitably reduced to acceptable values $\lesssim 10^{-4}-10^{-3}$ after
employing an iterative eccentricity-reduction procedure, thus leading to
accurate gravitational waveforms.

Finally, the evolution of the newly constructed ID has allowed us to obtain
a partial first estimate of the error budget introduced by the finite
resolution of the ID compared and to contrast it with the error
introduced by the resolution employed for the solution of the evolution
equations. While it is not in the scope of this paper to achieve a
complete quantitative analysis of the impact in the case of different initial
configurations, we have shown that the error budget contributed by the ID
resolution on the gravitational-wave phase evolution is in general
subdominant when compared to the errors introduced throughout the
evolution, even for relatively low ID resolutions. Of course, these
considerations only strictly apply to the configurations considered here and
to the resolutions employed both for the ID and the evolution, which are,
however, rather typical or real-life simulations of BHNS and BNSs.

Looking forward, there are multiple aspects of the spectral-solver
library presented here that can be improved in the future. First, the
current numerical setup inherits an assumed symmetry with respect to the
$(x,y)$ plane, so that only spinning configurations with spins aligned or
anti-aligned with the orbital angular momentum can be considered. There
are at least two different ways to further generalize this setup and thus
incorporate spin angular momenta that are not aligned along the
$z$-direction: either by generalising the domain decomposition and
relaxing the symmetry conditions enforced in the basis functions or by
splitting the tensor fields into symmetric and anti-symmetric
parts. Second, the system of equations is built and solved in the most
straightforward way possible, and often this is not necessarily
optimal. Considering that the Jacobian exhibits a structure that is known
a priori and that the latter is partly sparse, more efficient nonlinear
solvers could be employed, thus reducing the large memory demands and
computational costs of solving the system using the full Jacobian.
Third, the implementation of the black-hole boundary
conditions could be generalized to use locally a Kerr spacetime
background as done in Ref. \cite{Foucart2008}, thus enabling the solver
to cover the parameter space close to maximal black-hole spin angular
momentum.

The official release of the codes are available on the 
\texttt{Kadath} website:
\href{https://kadath.obspm.fr}{\texttt{https://kadath.obspm.fr}}.

\begin{acknowledgments}
ERM gratefully acknowledges support from a joint fellowship at the
Princeton Center for Theoretical Science, the Princeton Gravity
Initiative and the Institute for Advanced Study. The authors gratefully
acknowledge the Gauss Centre for Supercomputing e.V.
(www.gauss-centre.eu) for funding this project by providing computing
time on the GCS Supercomputer SuperMUC at Leibniz Supercomputing Centre
(www.lrz.de). Part of the simulations were performed on the national
supercomputer HPE Apollo Hawk at the High Performance Computing Center
Stuttgart (HLRS) under the grant number BBHDISKS. LR gratefully
acknowledges support from HGS-HIRe for FAIR and ``PHAROS'', COST Action
CA16214.
\end{acknowledgments}

\bibliographystyle{apsrev4-1}
%

\newpage
\appendix

\section{Eccentricity Reduction}
\label{sec:ecc}

To employ the eccentricity-reduction procedure mentioned extensively in
the main text, we have essentially utilised the methods detailed in
Refs. \cite{Buonanno2011, Foucart2008} as an effective manner to
iteratively reduce the eccentricity of our binary ID. In essence, once a
binary evolution is carried out and the distance between the two
components -- either a coordinate distance in the case of BBHs and BHNSs,
or a proper distance in the case of BNSs -- $r(t)$ and the corresponding
time derivative $\dot{r}(t)$ are measured and fitted using the following
ansatzes
\begin{align}
  r(t) &= S_r (t) - \frac{B_r}{\omega_r} \cos(\omega_r t + \phi_r)\,,
  && \label{equ:r-ecc-fit} \\
  \dot{r} (t) &= S_r (t) + B_r
  \sin(\omega_r t + \phi_r), &&
  \label{equ:dr-ecc-fit} 
\end{align}
where $B_r$, $\omega_r$, and $\phi_r$ are fitting parameters from which
it is possible to estimate the eccentricity $e$ as 
\begin{align}
  e := - \frac{B}{\omega_r d_0}\,,
\end{align}
with $d_0$ being the initial separation.

The function $S_r(t)$ in Eqs. \eqref{equ:r-ecc-fit} and
\eqref{equ:dr-ecc-fit} is freely specifiable and is used to fit, 
and hence remove any linear
regression in the data while the periodic term is used to
extract information regarding eccentricity induced oscillations in the orbit. 
In this work, we have used the following definition of $S_r(t)$ to produce
the results described in Sec. \ref{sec:res}
\begin{align*}
  S_r(t) &\equiv A_0 + A_1t\,. &&
\end{align*}
We have tested the use of quadratic terms in the expression for $S_r(t)$
and in the time dependencies of the oscillatory terms; however, this had
a negligible impact on the fit parameters for the binary configurations
considered in this work, hence prompting us to ignore these terms.

At each iteration, we use the fitting parameters $B_r$, $\omega_r$, and
$\phi_r$ to estimate the corrections, $\delta \dot{a}$ and $\delta \Omega$,
to $\dot{a}$ and $\Omega$ in Eq. \eqref{eq:betacor} using
\begin{align}
  \delta \dot{a} &:= - \frac{B \sin \phi}{d_0}, && \label{equ:adot:corr}\\
  \delta \Omega &:= - \frac{B \omega_r \cos \phi}{2 \Omega
    d_0}, && \label{equ:omega:corr}
\end{align}
so that the new shift in Eq. \eqref{eq:betacor} becomes
\begin{align}
  \label{eq:betacor_n}
  \beta^i_{{\rm cor}} = ( \Omega + \delta \Omega) 
  \partial^i_{\varphi}(\boldsymbol{x}_c) + (\dot{a} + \delta \dot{a}) r^i \,,
\end{align}
where the values of $\dot{a}$ and $\Omega$ are those obtained from the
previous iteration.

This procedure is iterated until the eccentricity is reduced to an
acceptable value, which, in all cases discussed in this work, was
obtained with four iterations. Note that fits using $r(t)$ provide
reasonable corrections until $e \approx 10^{-3}$. Attempting to reduce
eccentricity below this threshold required the use of $\dot{r}(t)$ as the
oscillations in $r(t)$ are too small to obtain an accurate fit.

\section{Post-Newtonian Estimates}
\label{sec:appendix_PN}

In the post-Newtonian framework, the equations of motion describing
circular motion in the center-of-mass frame corotating with the binary
become much simpler (see Ref. \cite{Blanchet2014}, Sec 7.4). In
particular, at the 3.5PN order, the quantities $\dot{a}$ and $\Omega$ can
be expressed as

\begin{widetext}
  \begin{align}
    \dot{a}_{3.5\rm{PN}} &= \frac{1}{r} \Bigg[ -\frac{64}{5} \frac{M^3 \nu}{r^3} \Big(1
      + \gamma \Big(-\frac{1751}{336} - \frac{7}{4} \nu \Big) \Big)
      \Bigg] \,,
    \label{equ:adot3PN}&& \\
    \Omega^2_{3.5\rm{PN}} &= \frac{M}{r^3} \Bigg[ 1 + \Big(-3 + \nu \Big) \gamma
      + \Big(6 + \frac{41}{4} \nu + \nu^2 \Big)\gamma^2
      + \Big(-10 - \frac{75707}{840} + \frac{41}{64} \pi^2 + 22 \ln\left(\frac{r}{r_0}\right) \nu
      + \frac{19}{2} \nu^2 + \nu^3 \Big) \gamma^3 \Bigg] \,.\label{equ:Omega3PN}&& 
  \end{align}
\end{widetext}
where $\mu := {M_1 M_2}/{M_\infty}$ is the reduced mass,$\nu:=q/(1+q)^2 = \mu / M_\infty$ is the symmetric mass
ratio, $r$ is the (coordinate) separation between the centres of the two
compact objects, and $r_0$ is the logarithmic barycentre defined by,
\begin{align}
	\ln r_0 &:= \frac{1}{M_\infty} \left(M_1 \ln r_1 + M_2 \ln r_2 \right)\,, &&
\end{align}
where $r_1$ and $r_2$ are the separation distances of the two compact
objects relative to the center-of-mass.

We have therefore used Eqs. \eqref{equ:adot3PN} and \eqref{equ:Omega3PN}
to obtain initial estimates for these quantities and employed them, for
instance, in the eccentricity-reduction procedure discussed in Appendix
\ref{sec:ecc}. Perhaps a bit unexpectedly, we have found that the ID
computed in this way provides a much better approximation to
quasi-circular orbits of more challenging configurations than ID obtained
assuming a quasi-equilibrium. This
is even more surprising since the approximations \eqref{equ:adot3PN} and
\eqref{equ:Omega3PN} do not take into account spin or spin-orbit
couplings. At the same time, it is important to underline that these
estimates do require an accurate measurement of the center-of-mass to
determine the radial position of each object relative to
center-of-mass. Therefore, quasi-equilibrium is an important initial
solution to obtain accurate PN estimates from Eqs. \eqref{equ:adot3PN}
and \eqref{equ:Omega3PN}. 

Additionally, Eqs. \eqref{equ:EB3PN} and \eqref{equ:EB4PN} have been used
to compute the binding energy curves shown in Figs. \ref{fig:bbh:binding}
and \ref{fig:bns:chi-eb}
\begin{widetext}
  \begin{align}
    E_{b, 3.5\rm{PN}} &= -\frac{\mu x}{2} \Bigg[ 1 + \Big(-\frac{3}{4} - \frac{1}{12} \nu \Big)x
      + \Big(-\frac{27}{8} + \frac{19}{8} \nu - \frac{1}{24} \nu^2 \Big)x^2
      + \Big(- \frac{675}{64} + \Big(\frac{34445}{576} - \frac{205}{96} \pi^2 \Big)\nu
      - \frac{155}{96} \nu^2 \nonumber \\
      &\phantom{= -\frac{\nu x}{2} \Bigg[}
        - \frac{35}{5184} \nu^3 \Big) x^3 \Bigg] \,, \label{equ:EB3PN} && \\
    E_{b,4\rm{PN}} &= E_{b,3.5\rm{PN}} + -\frac{\mu x}{2} \Bigg[ \Big(-
      \frac{3969}{128} + \frac{448}{15}\nu \ln x
      + e_4 \nu
      + \Big(- \frac{498449}{3456} + \frac{3157}{576} \pi^2 \Big) \nu^2
      +\frac{301}{1728} \nu^3
      +\frac{77}{31104} \nu^4 \Big) x^4 \Bigg] \,,\label{equ:EB4PN}&&
  \end{align}
  where $x:=\Omega^{{2}/{3}}$ and $e_4$ is the 4PN coefficient given by
  \begin{align}
    e_{4} := - \frac{123671}{5760} + \frac{9037}{1536} \pi^2 +
    \frac{1792}{15} \ln 2
    + \frac{896}{15} e \,.
  \end{align}
 
\end{widetext}

\end{document}